\documentclass[aps,prl,preprint,superscriptaddress]{revtex4}
\newcommand{\curl}{\nabla \times} % cf. confer
\usepackage{graphicx,subfigure,amssymb,amsmath,epsfig}

\begin{document}

%\preprint{}

\title{Validation of frequency and mode extraction calculations from time-domain
  simulations of accelerator cavities}

\author{Travis~M.~Austin}
\email{austin@txcorp.com}
\affiliation{Tech-X Corporation, 5621 Arapahoe Ave., Ste.~A, Boulder, CO 80303, USA}
\author{John~R.~Cary}
\affiliation{Tech-X Corporation, 5621 Arapahoe Ave., Ste.~A, Boulder, CO 80303, USA;
Center for Integrated Plasma Studies, University of Colorado, Boulder CO  80309, USA}
\author{Serguei~Ovtchinnikov}
\affiliation{Tech-X Corporation, 5621 Arapahoe Ave., Ste.~A, Boulder, CO 80303, USA}
\author{Gregory~R.~Werner}
\affiliation{Center for Integrated Plasma Studies, University of Colorado, Boulder CO  80309, USA}
\author{Leo~Bellantoni}
\affiliation{Fermi National Accelerator Laboratory, Batavia,
  IL, 60510, USA}

\date{\today}

\begin{abstract}
The recently developed frequency extraction algorithm
[G.~R.~Werner and J.~R.~Cary, J. Comp.~Phys.~227, 5200
(2008)] that enables a simple FDTD algorithm to be
transformed into an efficient eigenmode solver is applied to
a realistic accelerator cavity modeled with embedded
boundaries and Richardson extrapolation.  Previously, the
frequency extraction method was shown to be capable of
distinguishing $M$ degenerate modes by running $M$ different
simulations and to permit mode extraction with minimal
post-processing effort that only requires solving a small
eigenvalue problem.  Realistic calculations for an
accelerator cavity are presented in this work to establish
the validity of the method for realistic modeling scenarios
and to illustrate the complexities of the computational
validation process.  The method is found to be able to
extract the frequencies with error that is less than a part
in 10$^5$.  The corrected experimental and computed values
differ by about one parts in 10$^4$, which is accounted for
(in largest part) by machining errors.  The extraction of
frequencies and modes from accelerator cavities provides
engineers and physicists an understanding of potential
cavity performance as it depends on shape without incurring
manufacture and measurement costs.
\end{abstract}

\pacs{}
%\keywords{}

\maketitle

% main text
\section{Introduction}
\label{sec:back}

Once a computational application has been verified (shown to
correctly solve the mathematical model of a physical
system), for confidence in modeling, the application needs
to be validated, i.e., shown to accurately represent the
physical systems of interest.  Inevitably, validation is a
process of determining the causes of discrepancies between
computational and experimental results.  This paper focuses
on validation of frequency and mode extraction computations
of accelerator cavities using time-domain simulations.  The
calculations are based on the novel method of Werner and
Cary which employs time-domain simulations of Maxwell's
equations using the finite-difference time-domain (FDTD)
method \cite{WernerCary:2008}. The Werner-Cary method has
been shown to permit extraction of degenerate frequencies
and to provide the ability to reconstruct the modes using
only a minimal amount of post-processing linear algebra.

The common approach to frequency extraction for time-domain
simulations of Maxwell's equations using the FDTD method is
to compute narrowly-filtered states.  A narrowly-filtered
state can be computed using a square window that excites
only those modes around a desired frequency, $\omega$.  In
Maxwell's equations, this equates to the sinusoidal
excitation $f(t) = W(t) \sin(\omega t)$ where $W(t) =
\Theta(t)  \Theta(t-T)$ given that $\Theta(t)$ is the
Heaviside function.   One can also excite a broad range in a
similar way and then perform post-simulation FFTs on the
broadly-filtered states to determine the frequencies of the
system.  Such methods, however, do not provide a means for
constructing the eigenmodes or for handling degeneracies.

We apply the Werner-Cary method to a realistic accelerator
cavity that was fabricated at Fermi National Accelerator
Laboratory (Fermilab) in 1999.  The method is applied to
calculations performed with the VORPAL computational
framework \cite{Nieter:2004}.  Previous analysis of this
cavity shape had been carried out using the MAFIA code
(cf.~\cite{MAFIA:1999}), with stair-step and diagonal
boundaries and cylindrical symmetry
\cite{McAshan:2001,Bellantoni:2000}.  Accelerator cavities
are an essential component of large collider experiments
like the Large Hadron Collider.  The shape, curvature, and
size of these cavities determine the performance of the
cavities; simulations are important in the designing of
these cavities to maximize performance and minimize design
error.  We show the ability to get frequencies accurate to
less than a part in $10^5$.

The complexities of the validation process will be focused
on here as we illustrate the process that was undertaken to
determing the causes for discrepancies between computational
and experimental results.  Our requirement for validation is
that the discrepancies are reduced to measurement and/or
computational uncertainties. In the present case, we show
that the discrepancies are due to both computational error
and error associated with the cavity dimensions.  The latter
has two aspects.  The first is that cavities cannot be
fabricated precisely to specifications, and so one must do a
post-fabrication measurement of the cavity dimensions.  The
second is that even those post-fabrication measurements have
associated uncertainties.  For the particular case shown
here, it is found that the greatest uncertainties derive
from the uncertainties in cavity dimensions.  The
computational error is smaller by a few orders of magnitude.
This is not too surprising, in that accelerator cavities are
not primarily designed to have low frequency sensitivity
with respect to mechanical dimensions.  Nevertheless, this
shows how realistic frequency sensitivities are taken into
account in a validation study.

We begin with a review of the broadly-filtered
diagonalization method presented in
Ref.~\cite{WernerCary:2008}.  We then introduce the A15
accelerator cavity built at Fermilab. We describe the
details of the cavity and previous work on extracting
frequencies and modes of this cavity.  The subsequentsection
discusses our frequency computations from the specified
dimensions and their comparison with the measurements.  We
also consider resolution of the observed differences,
showing that dimensional uncertainties are dominant.  We
then present work on how modes can be obtained using the
Broadly-Filtered Filter Diagonalization Method.  We finally
summarize and conclude.

\section{Broadly-Filtered FDM \label{sec:ovrview}}

The Broadly-Filtered {\em Filter Diagonalization Method} (FDM) introduced in Ref.~\cite{WernerCary:2008} is an extension of the
concepts presented in Refs.~\cite{Neuhauser:1990,Neuhauser:1994,Wall:1995,Mandelshtam:1997,Mandelshtam:2003}.
We review the method in terms of linear operator ${\cal L}$ that represents a discrete $\curl \curl$ generated by the Yee method \cite{Yee:1966}.  
The field of interest, ${\bf s}$, is the elecric field, ${\bf E}$, or the magnetic field, ${\bf B}$.  If we view Maxwell's equation in the second-order form
\begin{equation}
\label{eq:redefine}
 - \frac{1}{c^2} \frac{\partial^2}{\partial t^2} {\bf E}(x,y,z,t) =
	\curl \curl {\bf E}(x,y,z,t) = {\cal L} {\bf E}(x,y,z,t),
\end{equation}
then the eigenvalue problem for $\curl \curl$ can be seen as a time evolution problem, relating eigenvalues to frequencies.

The broadly-filtered FDM is based on the idea that the linear operator, ${\cal L}$, can be transformed into block-diagonal form with
one large block filtered out and the remaining small block diagonalized by the singular value decomposition (SVD).  The
majority of the work goes into the diagonalization, which requires finding a small, invariant subspace of the field of
interest.  We generate this invariant subspace using a frequency filtering method based on targeted excitations of Maxwell's
equations using a current density that is geared towards the frequencies of interest.  See Eq.~(\ref{eq:specificCurrDens}).

The filtering is with respect to a FDTD discretization of Maxwell's equations for which the current source ${\bf J}(x,y,z,t)$ excites the 
electric and magnetic fields according to
\begin{equation}
\label{eq:1stOrdMaxwell}
\begin{array}{rcl}
\frac{\partial \mathbf{B}(x,y,z,t)}{\partial t}  & = & -\curl
\mathbf{E}(x,y,z,t) \\
\frac{1}{c^2} \frac{\partial \mathbf{E}(x,y,z,t)}{\partial t}  & = & \curl
\mathbf{B}(x,y,z,t) - \mu_0 \mathbf{J}(x,y,z,t).
\end{array}
\end{equation}
The current is defined as $\mathbf{J}(x,y,z,t) = f(t) \hat{\mathbf{J}}(x,y,z)$ where
\begin{equation}
\label{eq:currDens}
f(t) = \left\{
\begin{array}{lc}
2\left[ \frac{\sin(\omega_1(t - T/2))}{t - T/2} - \frac{\sin(\omega_2(t -
  T/2)) }{t - T/2} \right] \exp^{-\sigma_{\omega}^2 (t - T/2)^2/2} & 0 \le t \le T, \\
0. & \mathrm{otherwise}
\end{array}
\right.
\end{equation}
$\hat{\mathbf{J}}(x,y,z)$ has a pattern that encourages excitation of the desired modes in the frequency range $[\omega_1,\omega_2]$.  The parameter $\sigma_{\omega}$ is
determined by the separation of the frequencies in $[\omega_1,
\omega_2]$ from the next nearest frequency value.  If $\hat{\omega} < \omega_1$ is the nearest frequency, 
then  
\begin{equation}
\sigma_{\omega} < \frac{|\omega_1 - \hat{\omega}|}{5.68} 
\end{equation}
and
\begin{equation} 
T > \frac{11.4}{\sigma_{\omega}} 
\end{equation}
ensures that $\hat{\omega}$ and all other outside modes are suppressed by at least O($1e$-$7$).  Here, we have used the frequency amplitude of
a Gaussian-modulated sinusoid.  

If a $K$-degeneracy is known a priori to exist, then $K$ different spatial currents, $\hat{\mathbf{J}}$, are used  yielding $K$ simulations to extract the
$K$-degeneracy.  The $K$spatial currents should be chosen with an understanding of the symmetry of the cavity and the degenerate 
modes.

After the excitation is completed, the fields are temporally sampled at random grid points and then can be represented as a linear 
combination of the desired modes.   We then use this knowledge to extract the desired frequencies and spatial mode patterns.   To 
see this, consider the eigenvalue problem, ${\cal L} \mathbf{v}_m = \lambda_m
\mathbf{v}_m$, and the discrete fields, $\mathbf{s}_{\ell}$, sampled in time.  If we define
\begin{equation}
\label{eq:eigen}
\mathbf{r}_{\ell} = {\cal L} \mathbf{s}_{\ell}
\end{equation}
and use the fact that any eigenvector $\mathbf{v}_m$ can be expressed as a sum over the temporal samples according to
\begin{equation}
\label{eq:vm}
\mathbf{v}_m = \sum_{\ell=1}^L a_{m,\ell} \mathbf{s}_{\ell},
\end{equation}
then applying ${\cal L}$ yields
\begin{equation}
\label{eq:lmvm}
\lambda_m \mathbf{v}_m = \sum_{\ell=1}^L a_{m,\ell} \mathbf{r}_{\ell}.
\end{equation}
Since we have $\mathbf{s}_{\ell}$ and $\mathbf{r}_{\ell}$, Eqs.~(\ref{eq:vm}) and (\ref{eq:lmvm}) can be solved for the $\lambda_m$ (the eigenvalues) and the coefficients $a_{m,l}$
(allowing us to contruct the eigenvectors).

In practice, we only work with $P$ randomly-sampled spatial components of the fields yielding the generalized eigenvalue problem
\begin{equation}
\label{eq:geneig1}
R \mathbf{a}_m = \lambda_m S \mathbf{a}_m,
\end{equation}
where the number of colums of $R$ and $S$ correspond to the number of temporal samples and the number of rows correspond to
the number of spatial samples.  Eq.~(\ref{eq:geneig1}) can only be solved when $S$ is invertible, which requires $L=P$.  In general, we
prefer $P > 2L$ implying we must solve the generalized eigenvalue problem
\begin{equation}
\label{eq:geneig2}
S^{\dagger}  R \mathbf{a}_m = \lambda_m S^{\dagger}  S \mathbf{a}_m,
\end{equation}
using the SVD for $S^{\dagger} S = V D^2 V^{\dagger}$, where $D$is diagonal and $V$ is orthogonal.  Since $S^{\dagger} S$ may be singular, $D$ may have zeros on the 
diagonal, so we construct $\tilde{D}$ such that
\[
\tilde{D}^{-1}_{\ell \ell} = \left\{
\begin{array}{cl}
D^{-1}_{\ell \ell} & \frac{D_{\ell \ell} }{D_{\mathrm{max}}} >
  \alpha_{\mathrm{cutoff}} \\
0 & \frac{D_{\ell \ell} }{D_{\mathrm{max}}} \le
  \alpha_{\mathrm{cutoff}}
\end{array}
\right.
\]
where $\alpha_{\mathrm{cutoff}}$ is a small value chosen to distinguish significant diagonal values from the insignificant diagonal values.  We then solve 
the eigenvalue problem
\begin{equation}
\label{eq:geneig3}
V (\tilde{D})^{-2} V^{\dagger} S^{\dagger} \, R \mathbf{a}_m = \lambda_m \mathbf{a}_m,
\end{equation}
computing the eigenvalues $\lambda_m$.   The coefficients, $\mathbf{a}_m$, are then used to construct the field pattern, i.e., the eigenmode, by using 
Eq.~(\ref{eq:vm}).  We refer the reader to Ref.~\cite{WernerCary:2008} for further details.

%%%%%%%%%%%%%%%%%%%%%%%%%%%%%%%%%%%%%%%%%%%%%%%%%%%%%%%%%%%%%%%%%%%%%%%
%  FIGURE - Commented out for journal and submitted separately
%    File : surfview.eps  endplateview.eps
\begin{figure}[t]
\begin{center}
\subfigure[$\, $Side View]{\epsfig{figure=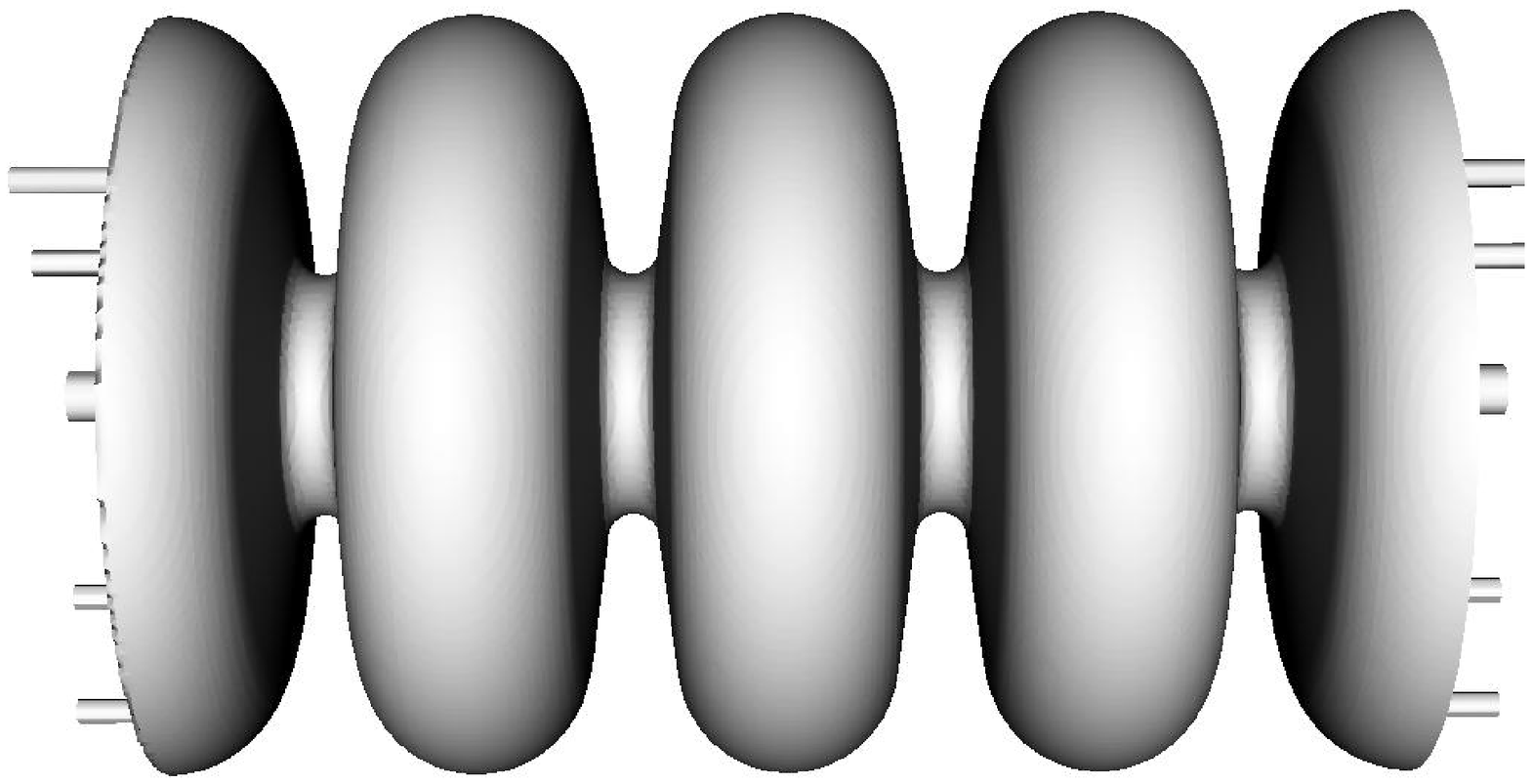,width=65mm}} % figures/surfview.png
\hspace{10mm}
\subfigure[$\, $End-on View]{\epsfig{figure=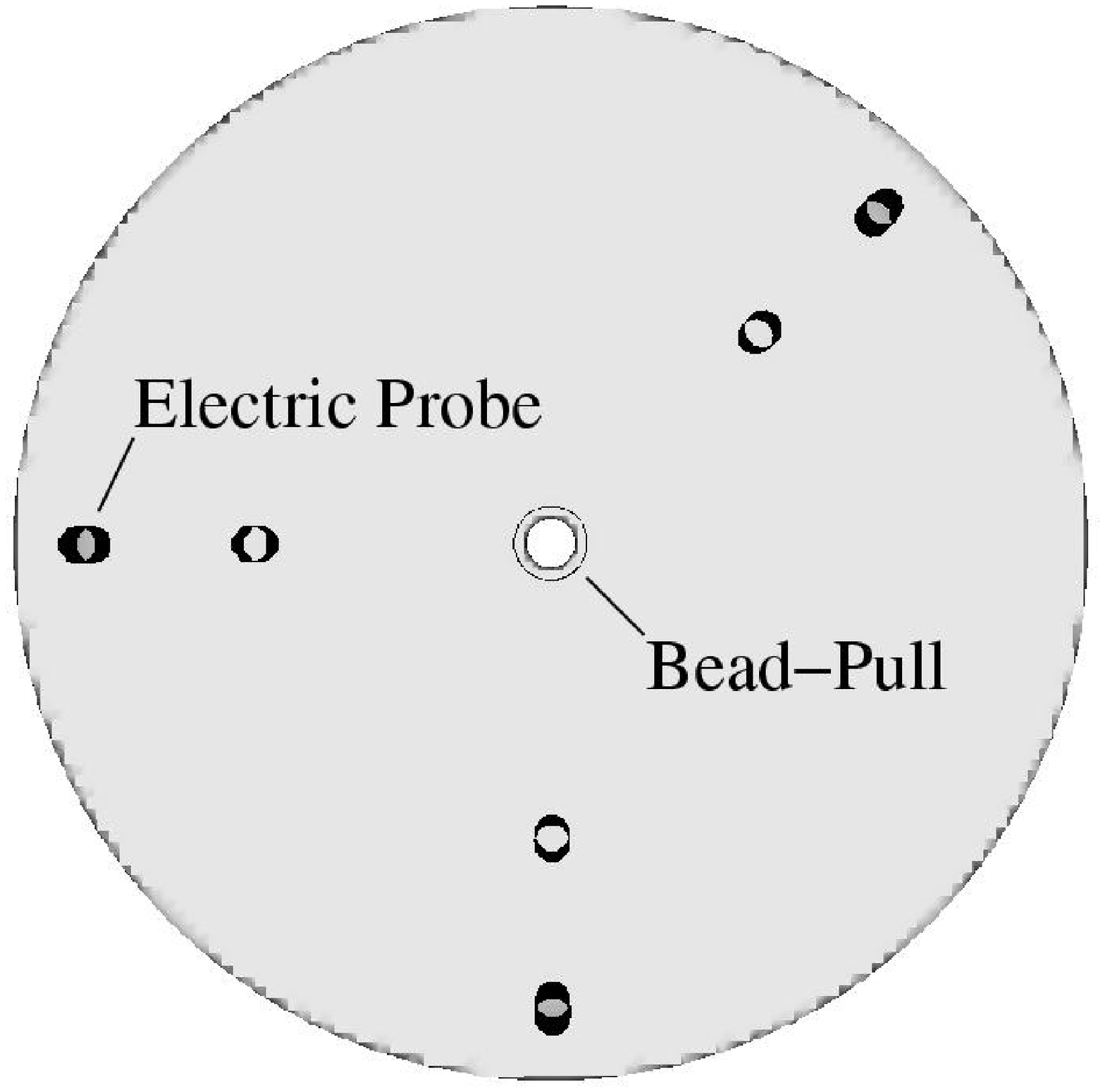,width=40mm}} % figures/endplateview.png
\caption{A Side (a) and End-on (b) view of the A15 cavity. The
  center hole in the End-on view labeled {\em Bead-Pull} was used for
  bead-pull experiments and the smaller off-center
  endplate hole labeled {\em Electric Probe} held an electric probe that created
  a dipole perpendicular to the page. \label{fig:a15surfview}}
\end{center}
\end{figure}
%%%%%%%%%%%%%%%%%%%%%%%%%%%%%%%%%%%%%%%%%%%%%%%%%%%%%%%%%%%%%%%%%%%%%%%

\section{A15 Cavity}
\label{sec:a15}

In Ref.~\cite{WernerCary:2008} thirteen TM modes in a rectangular 2D cavity were found using the frequency extraction algorithm.  The 
standard Yee method was used to discretize Maxwell's equations and to evolve the electric and magnetic fields \cite{Yee:1966}.  Here, 
validation is performed for an extensively tested stack of four unpolarized dumbbells from Fermilab.  We refer to this stack, which is
pictured in Fig.~\ref{fig:a15surfview}, as the A15 cavity.

\begin{figure}[bh]
\begin{center}
\epsfig{figure=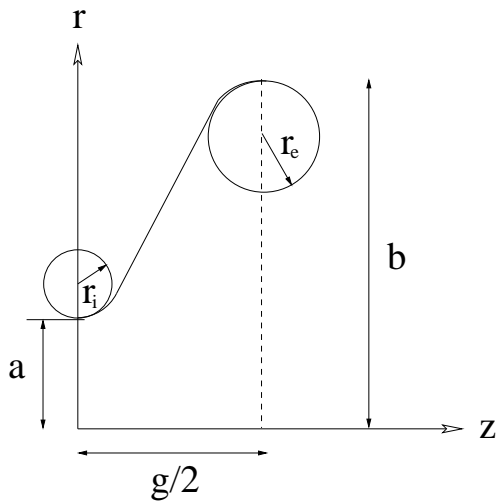,width=65mm} % figures/cavityshapescheme.png
\caption{Schematic of the A15 cavity describing the shape in relation
  to the equatorial radius ($b$), the iris radius ($a$), the iris curvature
  ($r_i$), the equatorial curvature ($r_e$), and the cell half length
  ($g/2)$\label{fig:a15scheme}}
\end{center}
\end{figure}

The A15 cavity was designed in 1999 for the development of a separated K$^+$ beam \cite{Bellantoni:2001}.   It is a deflecting
mode cavity designed to operate at 3.9 GHz, i.e., the $\pi$ mode.  The cavity shape is determined by parameters such as the equatorial 
radius ($b$), iris radius ($a$), iris curvature ($r_i$), equatorial curvature ($r_e$), and cell half length ($g/2)$, displayed schematically in
Fig.~\ref{fig:a15scheme}.  For the A15 cavity, these parameters are 47.19 mm, 15.00 mm, 3.31 mm, 13.6 mm, and 19.2 mm.

%%%%%%%%%%%%%%%%%%%%%%%%%%%%%%%%%%%%%%%%%%%%%%%%%%%%%%%%%%%%%%%%%%%%%%%
%  FIGURE - Commented out for journal and submitted separately
%    File : cavityshapescheme.eps  endplateview.eps
%%%%%%%%%%%%%%%%%%%%%%%%%%%%%%%%%%%%%%%%%%%%%%%%%%%%%%%%%%%%%%%%%%%%%%%

The endplate holes observed in the end-on view of Fig.~\ref{fig:a15surfview} were used for experimental purposes.  The large center hole 
positioned on the centerline of the cavity (radius=3.175 mm) was used for bead-pull experiments.  One of the off-center holes
(radius=1.5875 mm) provided space for an electric probe that created a dipole.  Furthermore, an equivalent set of holes exists on the 
other endplate.  Cavities used within working accelerators would be without these holes; however, these holes only have a small
effect on frequencies (tens of kHz) and validation.

In Ref.~\cite{Bellantoni:2000} the frequencies of the A15 cavity were reported from bead-pull experiments.  The frequencies were
corrected for temperature, barometic pressure, and relative humidity to the frequency that would be measured in a vacuum at
25$^{\circ}$ C yielding the values presented in Table~\ref{tab:deflectFreqs}.   In addition to the TM$_{110}$ modes, Ref.~\cite{Bellantoni:2000} also 
had frequencies for the TM$_{010}$ modes as well as other higher-order modes.

\begin{table}[tb]
\begin{center}
\begin{tabular}{|c|c|c|c|c|c|} \hline
 & $f_0$ & $f_1$ & $f_2$ & $f_3$ & $f_4$ \\
\hline
Exact & 3902.810 & 3910.404 & 3939.336 & 4001.342 & 4106.164 \\
\hline 
Computed & 3900.512 & 3908.552 & 3937.325 & 4000.107 & 4103.616 \\
\hline
\end{tabular}
\end{center} \caption{\label{tab:deflectFreqs} The first line is the set of frequencies
(in MHz) of the five deflecting modes for the A15 cavity.  Measurements were performed 
at Fermi National Accelerator Laboratory in 2000 by Bellantoni {\em et al.} 
and presented in the unpublished note \cite{Bellantoni:2000}.  The second line is the
initial set of computations using the original specifications in \cite{Bellantoni:2000} and
discussed in the section {\em Frequency Calculations}.}
\end{table}

Applying the broadly-filtered FDM to the A15 cavity requires boundary methods that approximate Maxwell's equations on the
cells cut by the cavity boundary.  We used the Dey-Mittra method \cite{dey1997lcf} for solving Maxwell's equations on the cavity's
curved boundaries.  The Dey-Mittra method requires that one exclude small fractional areas, with the exclusion depending on the 
ratio $f_{DM} \equiv \Delta t/\Delta t_{CFL}$ of simulation time step to the CFL time step of a system without embedded boundaries.  A method for 
choosing which areas are excluded that guarantees numerical stability is discussed in \cite{Nieter:2009}.  For our simulations, we 
used $f_{DM} = 0.1$.

\section{Frequency Calculations}

Previously, we described how the frequency filtering approach for Maxwell's equation involved selecting an appropriate time-varying
current that selectively excites a given frequency range. Defining $f_L =$ 3902 MHz and $f_U =$ 4110 MHz, the frequency filtering aproach 
requires knowledge of the separation between $[f_L,f_U]$ and the nearby modes.   From Ref.~\cite{Bellantoni:2000} it was known that 
the mode nearest to the range was the mode at $f_b =$ 4320 MHz.   When previous calculations or experiments are not available, a few quick
simulations followed by FFTs can identify rough locations of modes.

To reduce this nearest unwanted mode by $h = 1e$-$7$, we must run the excitation for at least 200 oscillations at $4000$ MHz, since
\begin{equation}
\label{eq:freqEsti}
 f T = \frac{32}{\pi}
\frac{f}{|f_b- f_{U}|}.
\end{equation}
See Ref.~\cite{WernerCary:2008} for details.  This suppresses modes more than 200 MHz away from $f_U$, i.e., frequencies above 4310 MHz. 
Adding the time for sampling of $\pi/|f_b - f_{U}|$ the total number of oscillations is approximately 260 oscillations.  

We have also judiciously chosen the spatial pattern, $\hat{\mathbf{J}}(x,y,z)$, using a priori knowledge of the modes of interest.  When there is no 
known spatial pattern, the user can create a random pattern  for $\hat{\mathbf{J}}$.  For the A15 cavity, we used the spatial pattern given by
\begin{align}
\label{eq:specificCurrDens}
\hat{J}_x(x,y,z) := j_T(y,z) \, [ a_1 & \cos (2 k x) + a_2
\cos(4 k x) +  b_1 \cos(k x) \\
  &  + b_2 \sin(3 k x)  + b_3 \sin(k x) ] \nonumber
\end{align}
where $k=\pi/0.1536$, $j_T(y,z) = y$ or $z$, and $\hat{\mathbf{J}}
= (\hat{J}_x,\hat{J}_y,\hat{J}_z )^T$.  The coefficients $a_{m}$ and $b_{n}$ are chosen at random for a given 
simulation, thus running multiple simulations to properly account for the two polarizations split by only a few kHz at each frequency 
value just requires that we have random coefficients for each simulation.

We computed the frequencies for different cell sizes.  In all cases, we used $\Delta y = \Delta z = 1.25 \Delta x$, where $x$ denotes the longitudinal 
coordinate.  The cell size $\Delta x$varied from 0.533 mm to 0.267 mm implying that the cell sizes, $\Delta y$, and $\Delta z$, vary from 0.666 mm to 
0.333 mm, thus resolving the hole sizes of 3.175 mm and 1.5875 mm.  At the given resolutions and the given Dey-Mittra 
fractional face parameter, it was observed that the Yee method combined with the Dey-Mittra boundary algorithm was second-order.

Since the method was second-order, Richardson extrapolation could be used to achieve a third-order method, i.e., one makes the 
assumption,
\begin{equation}
\label{richardson}
\omega_i \sim \omega_0 + \alpha/N_i^2,
\end{equation}
where $\omega_i$ is the $i^{th}$ computed frequency, $\omega_0$ is the true frequency, $N_i$ is the number of cells across the simulations in any particular 
direction (given that all directions are refined together) for that computation, and $\alpha$ is the error coefficient.  Using two computed 
frequencies at two different resolutions, one gets two equations generated from Eq.~(\ref{richardson}) for both $\omega_0$ and $\alpha$.  Given more 
than two points, one can also determine the frequency from a regression analysis.  Using these multiple means of extraction to compute 
the frequency and then finding the standard deviation of the mean of those results, we were able to deduce that the computational error 
was $O(10-40)$ kHz for the range of frequencies.  That is, the modes can be found to an accuracy that is less than a part in $10^5$.

\begin{figure}[bht]
\begin{center}
\subfigure{\epsfig{figure=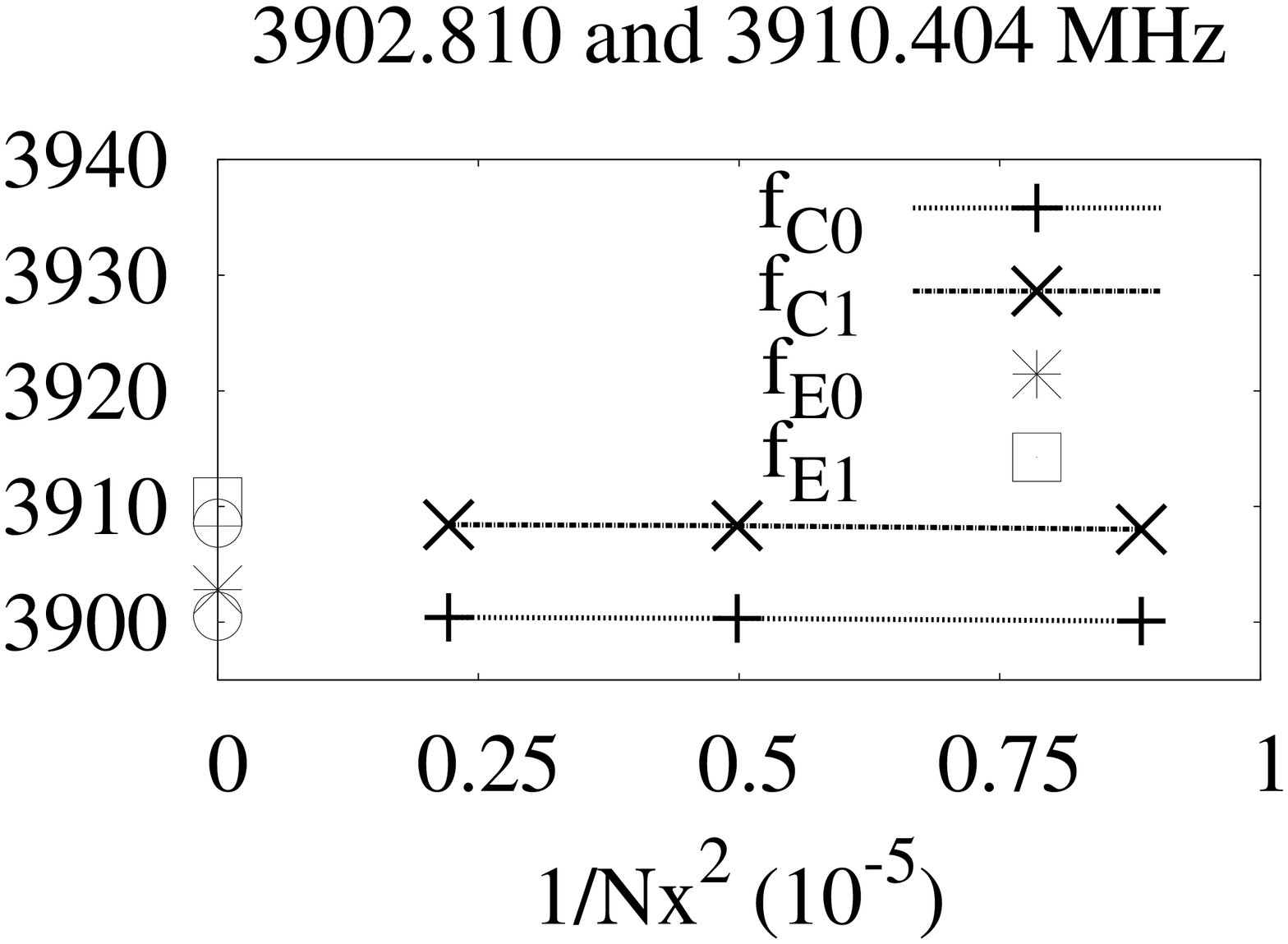,width=72mm}} % gnuplot/a15cav2conv1.pdf
\hspace{-10mm}
\subfigure{\epsfig{figure=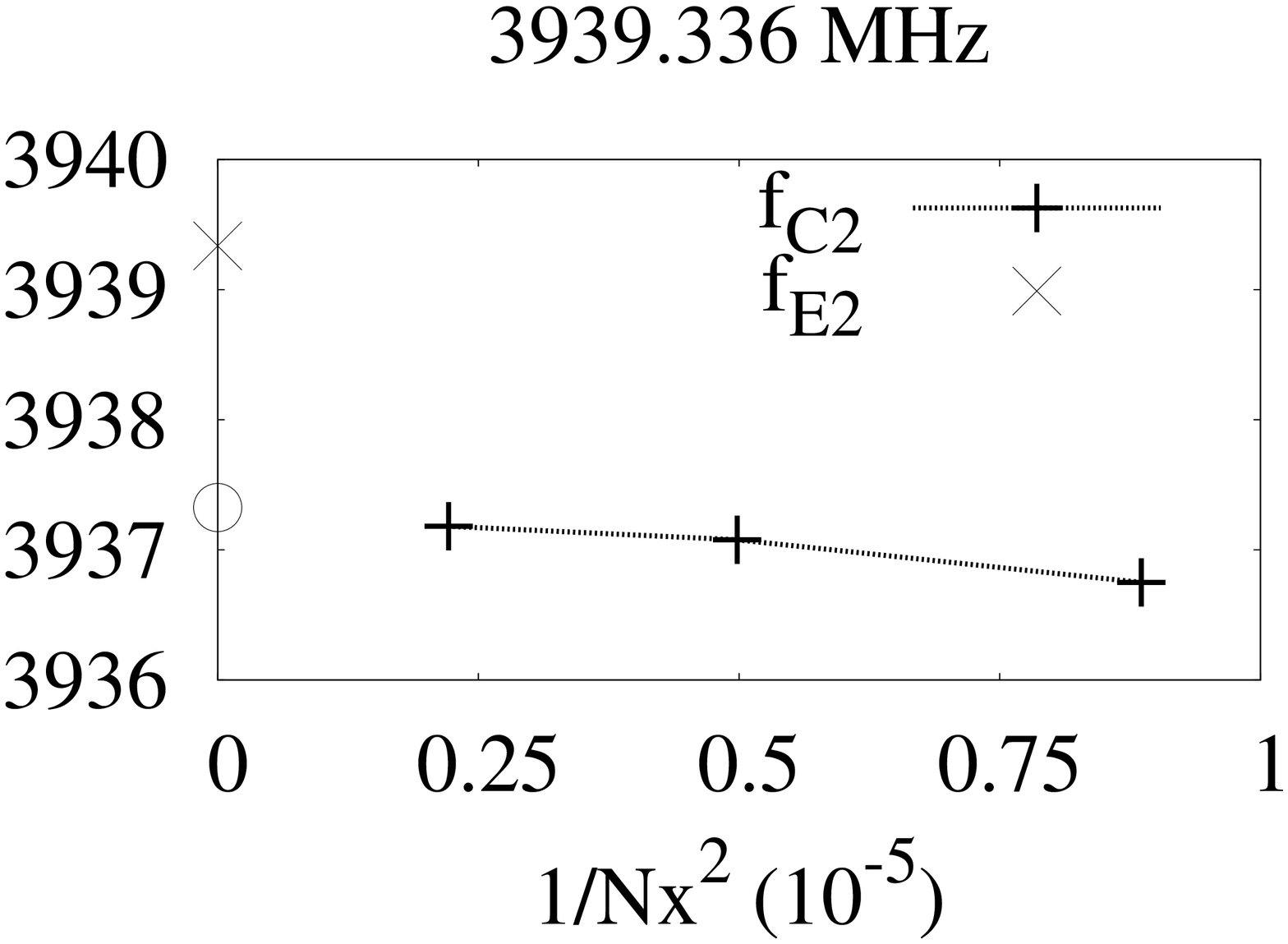,width=72mm}} % gnuplot/a15cav2conv2.pdf
\\
\subfigure{\epsfig{figure=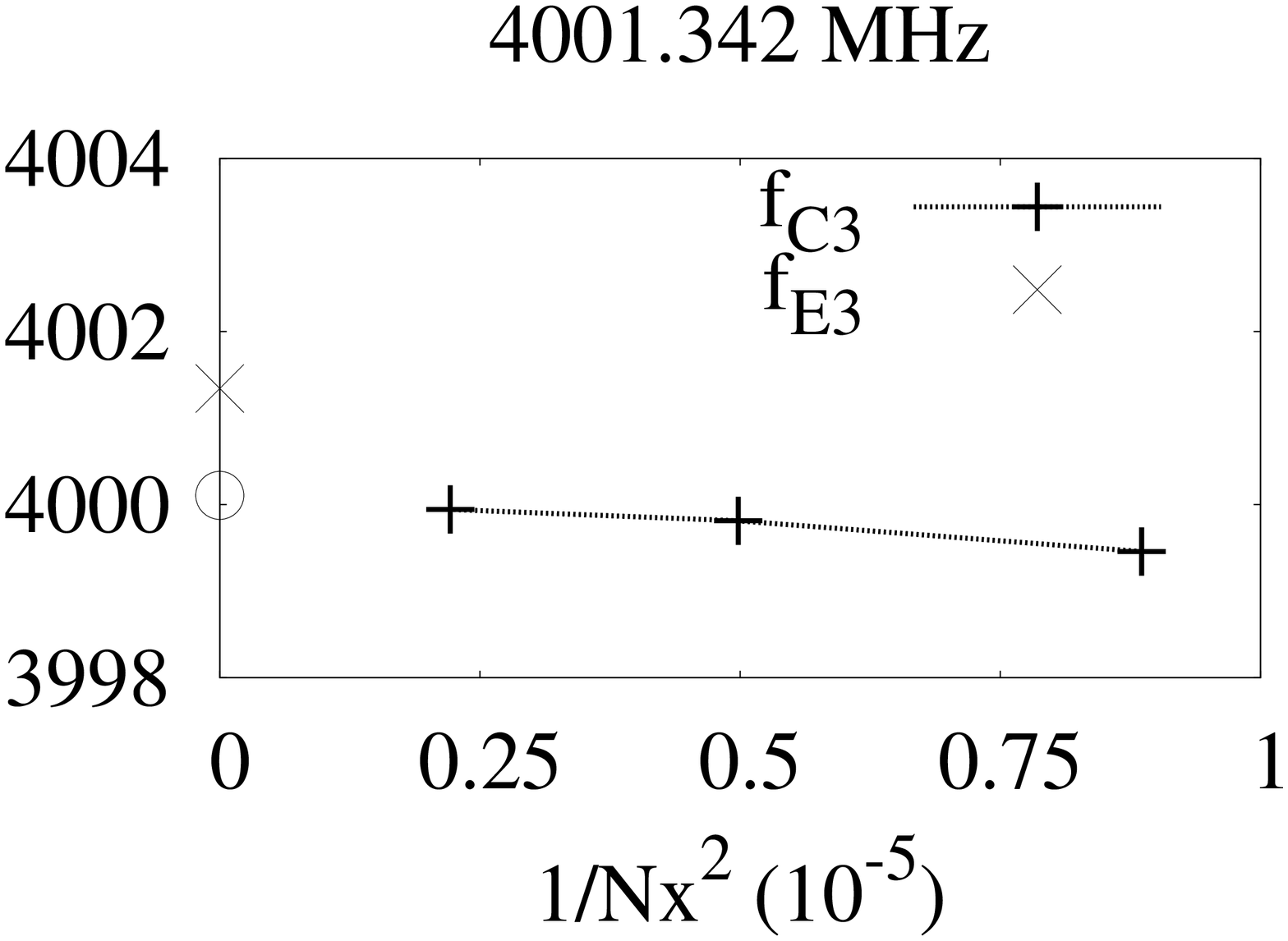,width=72mm}} % gnuplot/a15cav2conv3.pdf
\hspace{-10mm}
\subfigure{\epsfig{figure=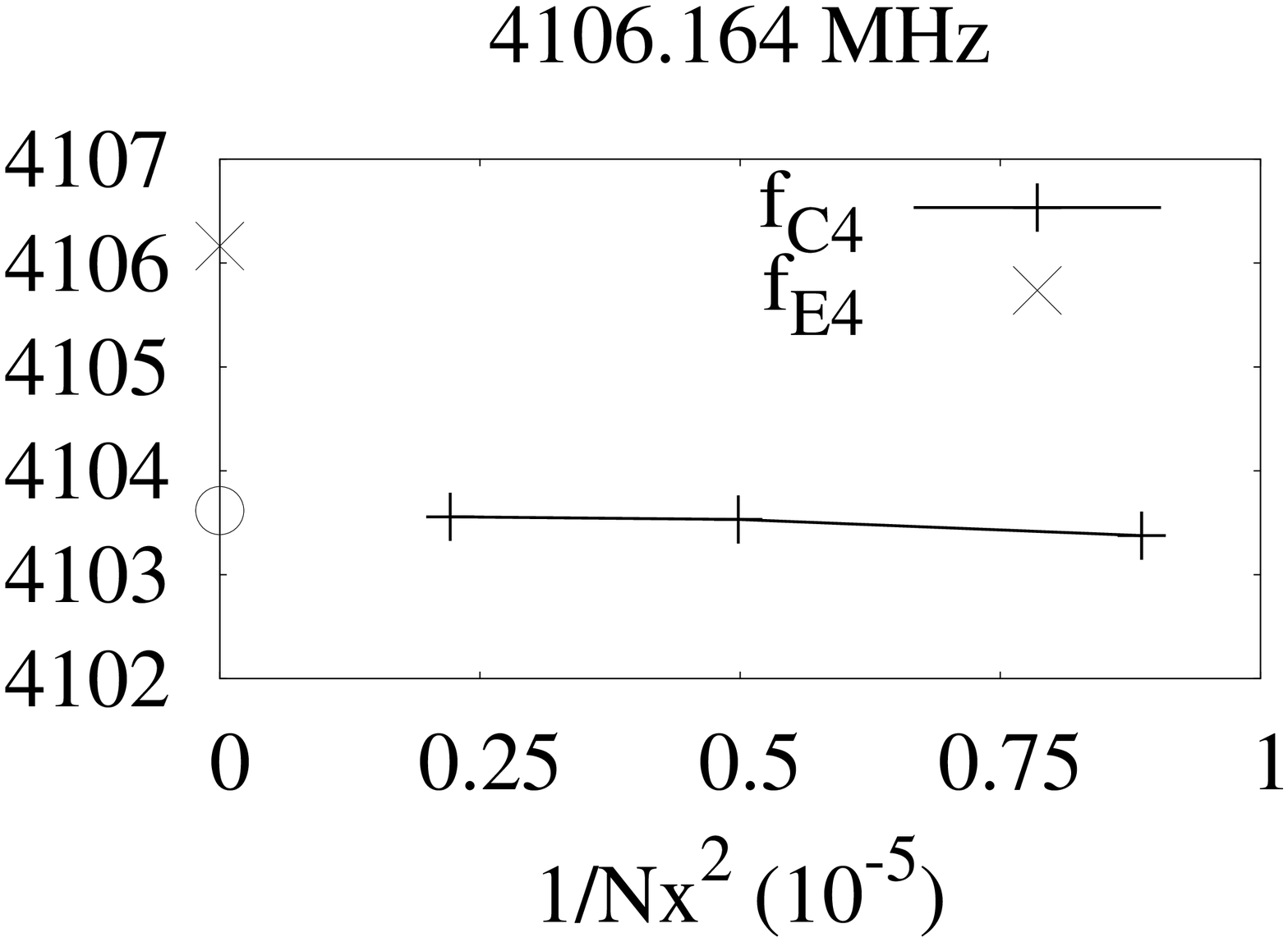,width=72mm}} % gnuplot/a15cav2conv4.pdf
\caption{Convergence plot (f$_{\mathrm{C}N}$) of the frequencies of
  the five deflecting modes for the cavity with the dimensions
  specified in Ref.~\cite{Bellantoni:2000}.  The abscissa
  represents the grid resolution and the ordinate corresponds to the
  frequency value in MHz.  Each plot has on the ordinate the experimental
  values (f$_{\mathrm{E}N}$) from Table \ref{tab:deflectFreqs} for
  reference purposes and the Richardon's extrapolated values as unfilled
  circles.
\label{fig:a15cav2conv}}
\end{center}
\end{figure}

In Fig.~\ref{fig:a15cav2conv}, we have plotted the convergence of computed frequencies and the value obtained with Richardson
extrapolation as an unfilled circle on the ordinate axis.  Also shown on the ordinate axis are the experimental values presented
in Ref.~\cite{Bellantoni:2000}.   Note that there are two polarizations at each frequency value with a slight split in the values.  These values
differ by anywhere from 3-6 kHz.  An averaged value has been used in the plots and in the Richardson extrapolated values.  

\begin{figure}[bht]
\begin{center}
\subfigure{\epsfig{figure=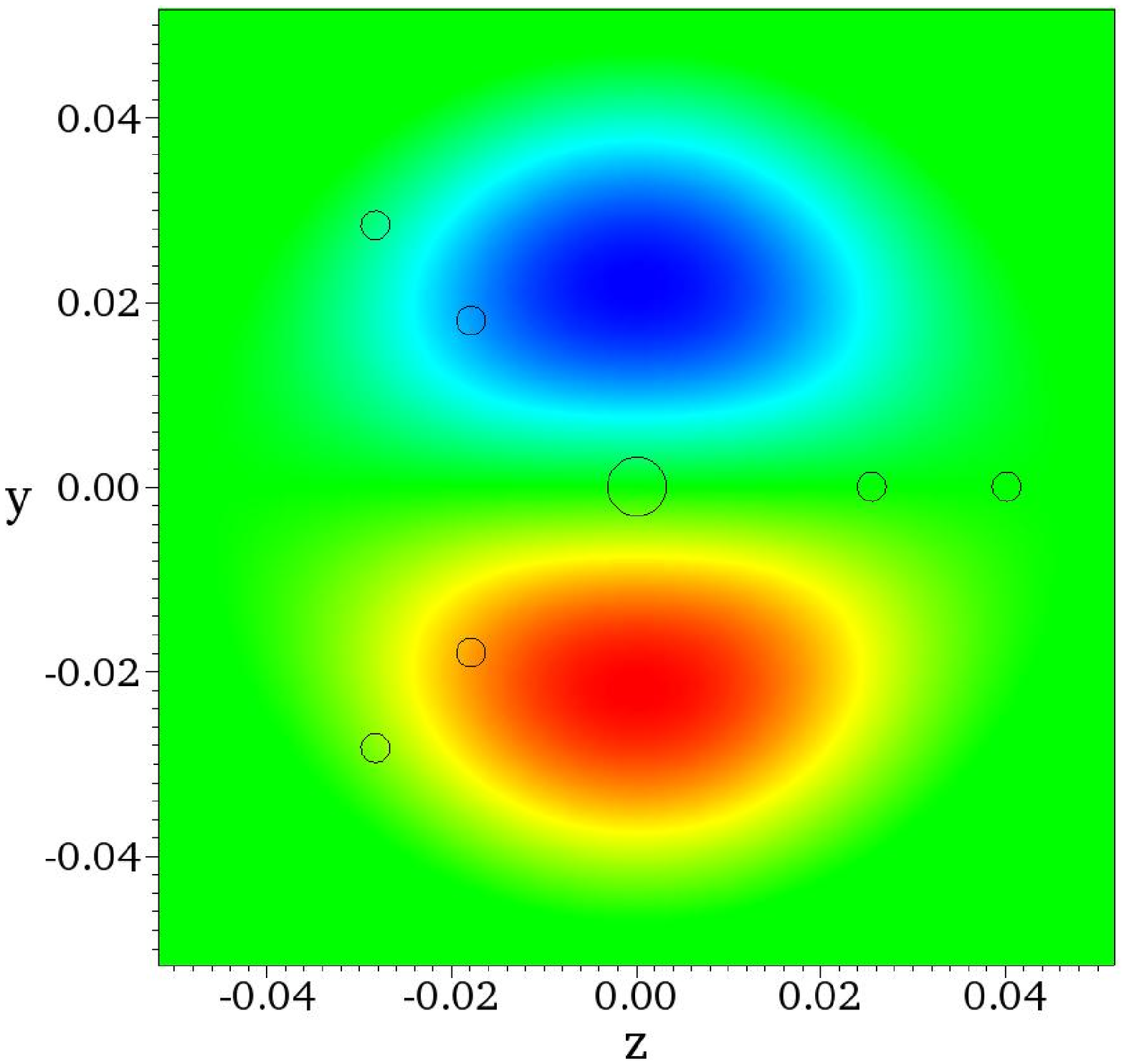,width=50mm}} % Visit/mode0b_0pi.png
\hspace{1mm}
\subfigure{\epsfig{figure=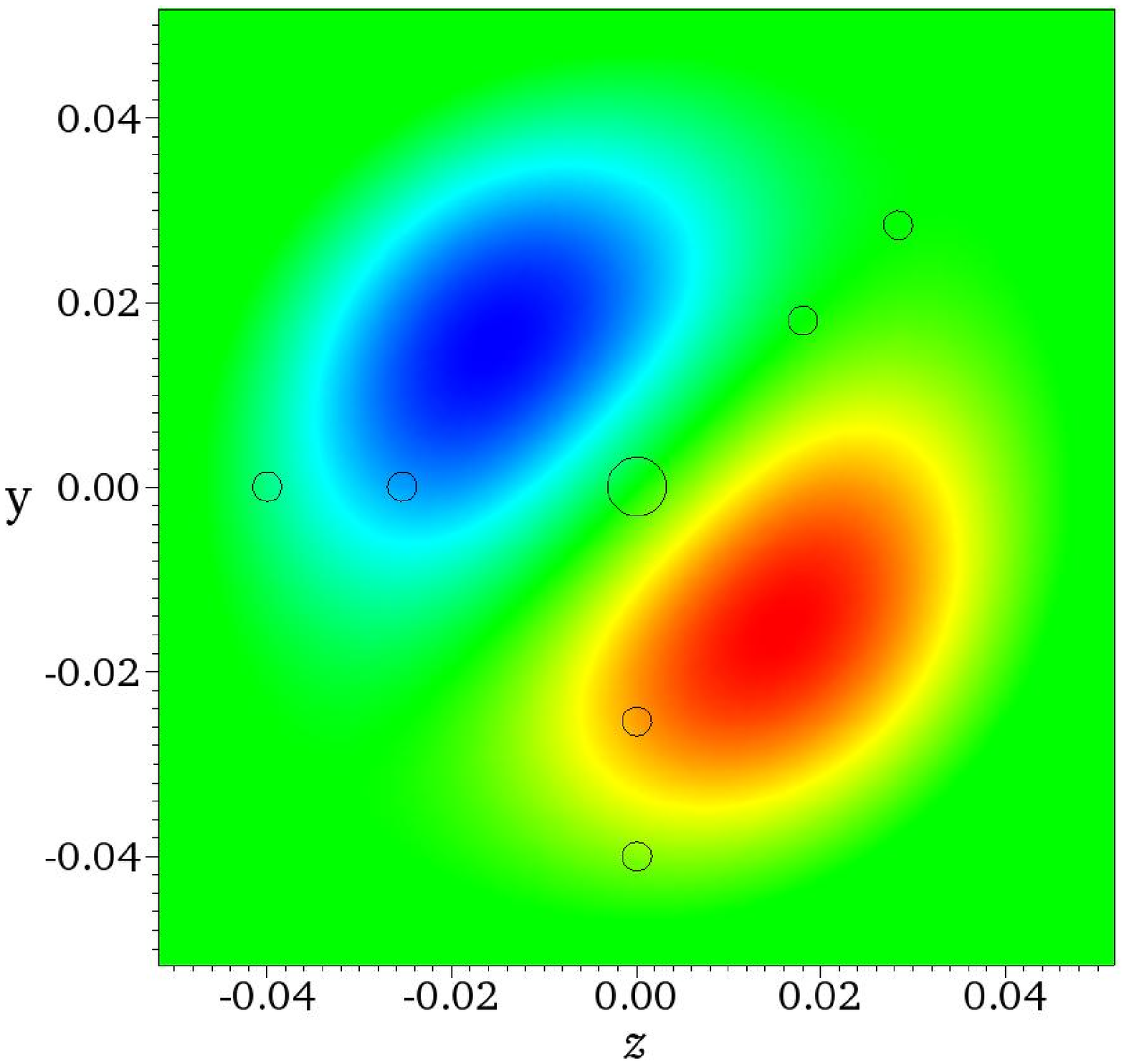,width=50mm}} % Visit/mode0b_pi2.png
\hspace{1mm}
\subfigure{\epsfig{figure=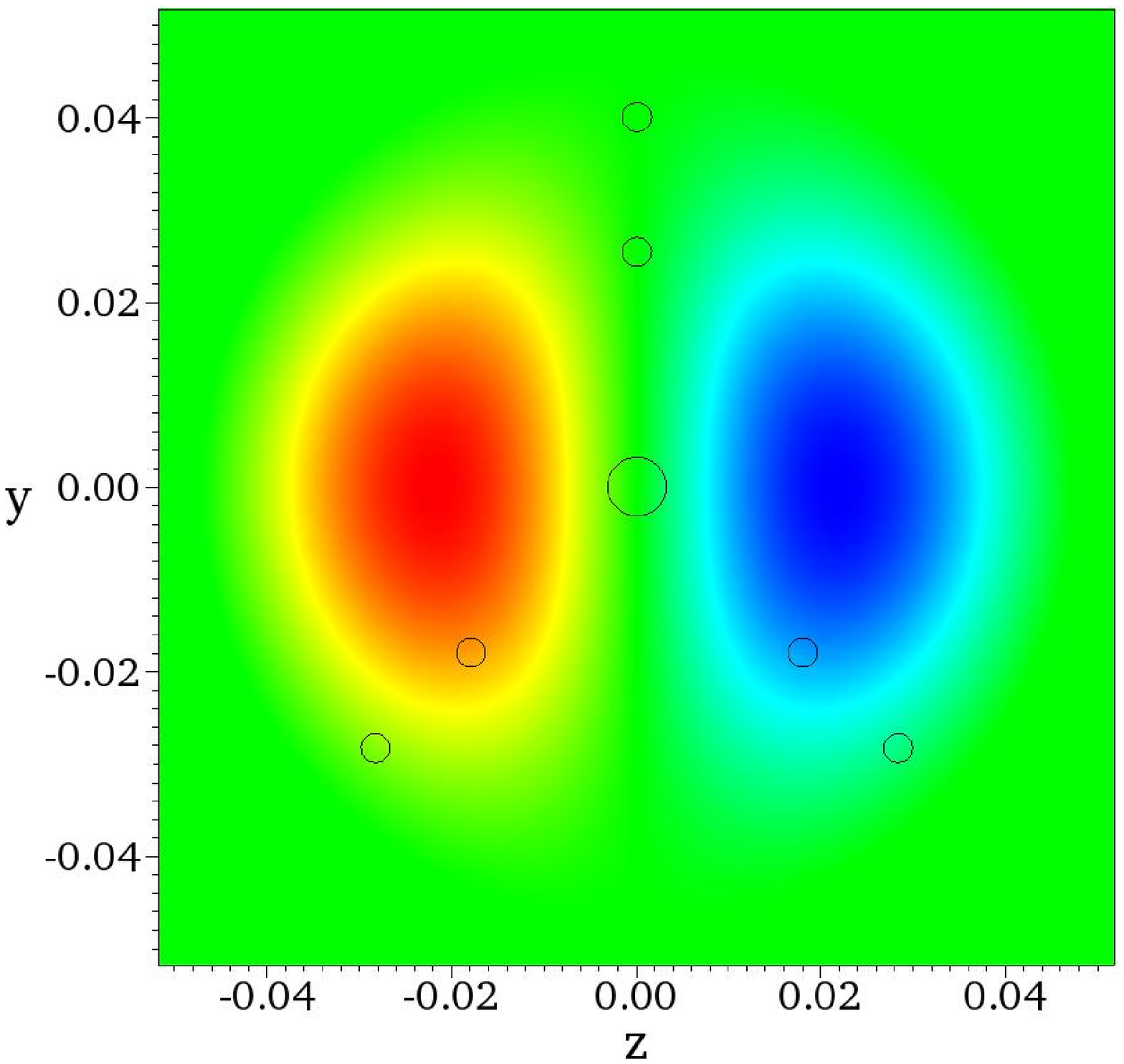,width=50mm}} % Visit/mode0b_pi4.png
\caption{Slice of the $E_x$ component for one of the two polarizations of the $\pi$ mode for the various rotations
in the computational domain of the cavity.  The other polarization was orthogonal to the one shown and the difference 
in frequency between the two polarizations was only a few kHz.  Also to be noted is the arrangement of the polarization 
due to the endplate hole alignment.
\label{fig:a15cav2_dipole_allholes}}
\end{center}
\end{figure}

In Fig.~\ref{fig:a15cav2_dipole_allholes}, we have plotted a slice of the $E_x$ component in the $yz$-plane for one of the two polarizations 
of the $\pi$ mode.  We have produced this plot for various rotations about the longitudinal direction of the cavity in the computational 
domain the method's ability to capture the dependency of the splitting of the polarizations on the endplate holes and not the grid.  The 
polarization that is not shown was orthogonal to the one in the figure.   When no endplate holes are used, the two polarizations are still 
orthogonal but their configuration is based on the transverse components of the currents used to excite the modes, i.e.~Eq.~(\ref{eq:specificCurrDens}).

%%%%%%%%%%%%%%%%%%%%%%%%%%%%%%%%%%%%%%%%%%%%%%%%%%%%%%%%%%%%%%%%%%%%%%%
%  FIGURE - Commented out for journal and submitted separately
%   File : a15cav2conv1.eps a15cav2conv2.eps
%          a15cav2conv3.eps a15cav2conv4.eps
%%%%%%%%%%%%%%%%%%%%%%%%%%%%%%%%%%%%%%%%%%%%%%%%%%%%%%%%%%%%%%%%%%%%%%%

From Fig.~\ref{fig:a15cav2conv} the discrepancies between the computed and measured values are apparent.  The values obtained with 
Richardson extrapolation are also presented numerically in Table~\ref{tab:deflectFreqs}.  They are seen to differ from the experimental 
measurements in Table~\ref{tab:deflectFreqs} by 1.5--2.5 MHz.  This difference significantly exceeds the computational uncertainty.  
Hence, we must understand the origin of the difference.

\section{Validation: resolution of differences}

The differences (1.5--2.5 MHz) between the computed values and the experimentally measured values of Table \ref{tab:deflectFreqs} 
exceed the estimated computational uncertainty; in this section we describe how we resolved these differences.  We began by considering 
possible sources of the discrepancy: computational error, missing physics in the simulation, and cavity (geometry) measurement error.

%%%%%%%%%%%%%%%%%%%%%%%%%%%%%%%%%%%%%%%%%%%%%%%%%%%%%%%%%%%%%%%%%%%%%%%
%  FIGURE - Commented out for journal and submitted separately
%   File : a15cav3conv1.eps a15cav3conv2.eps
%          a15cav3conv3.eps a15cav3conv4.eps
\begin{figure}[t]
\begin{center}
\subfigure{\epsfig{figure=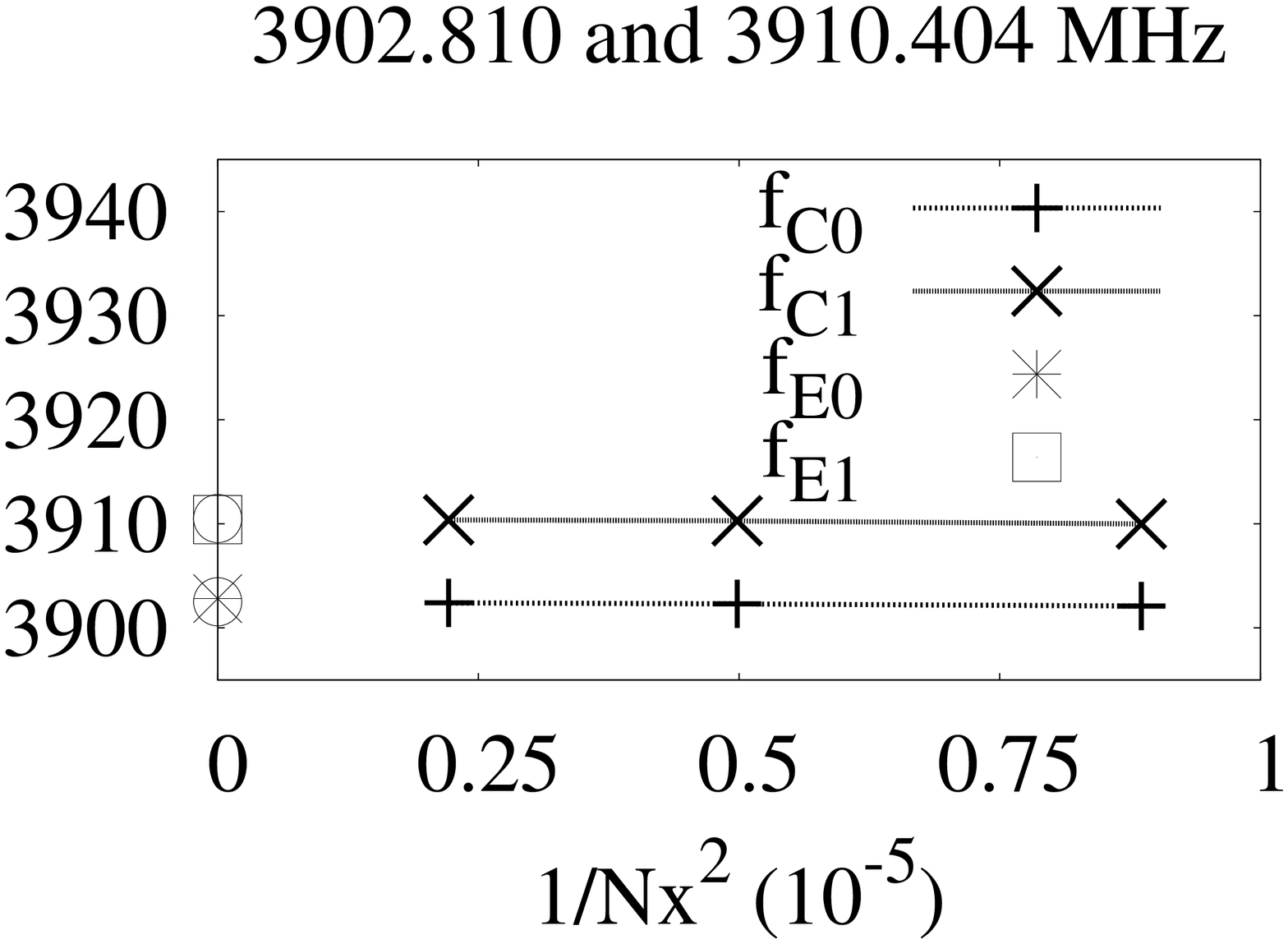,width=72mm}} % gnuplot/a15cav3conv1.pdf
\hspace{-10mm}
\subfigure{\epsfig{figure=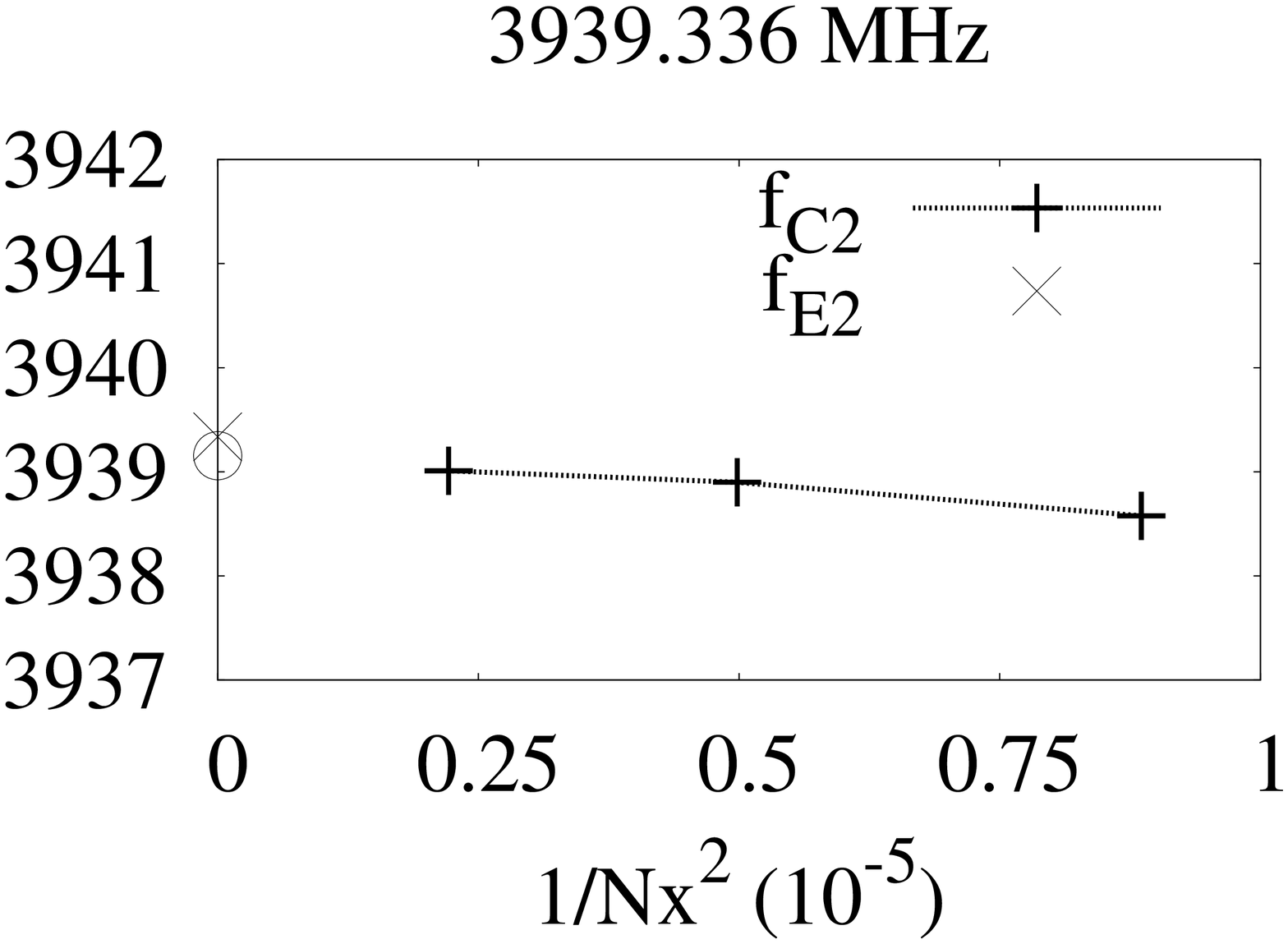,width=72mm}} % gnuplot/a15cav3conv2.pdf
\\ 
\subfigure{\epsfig{figure=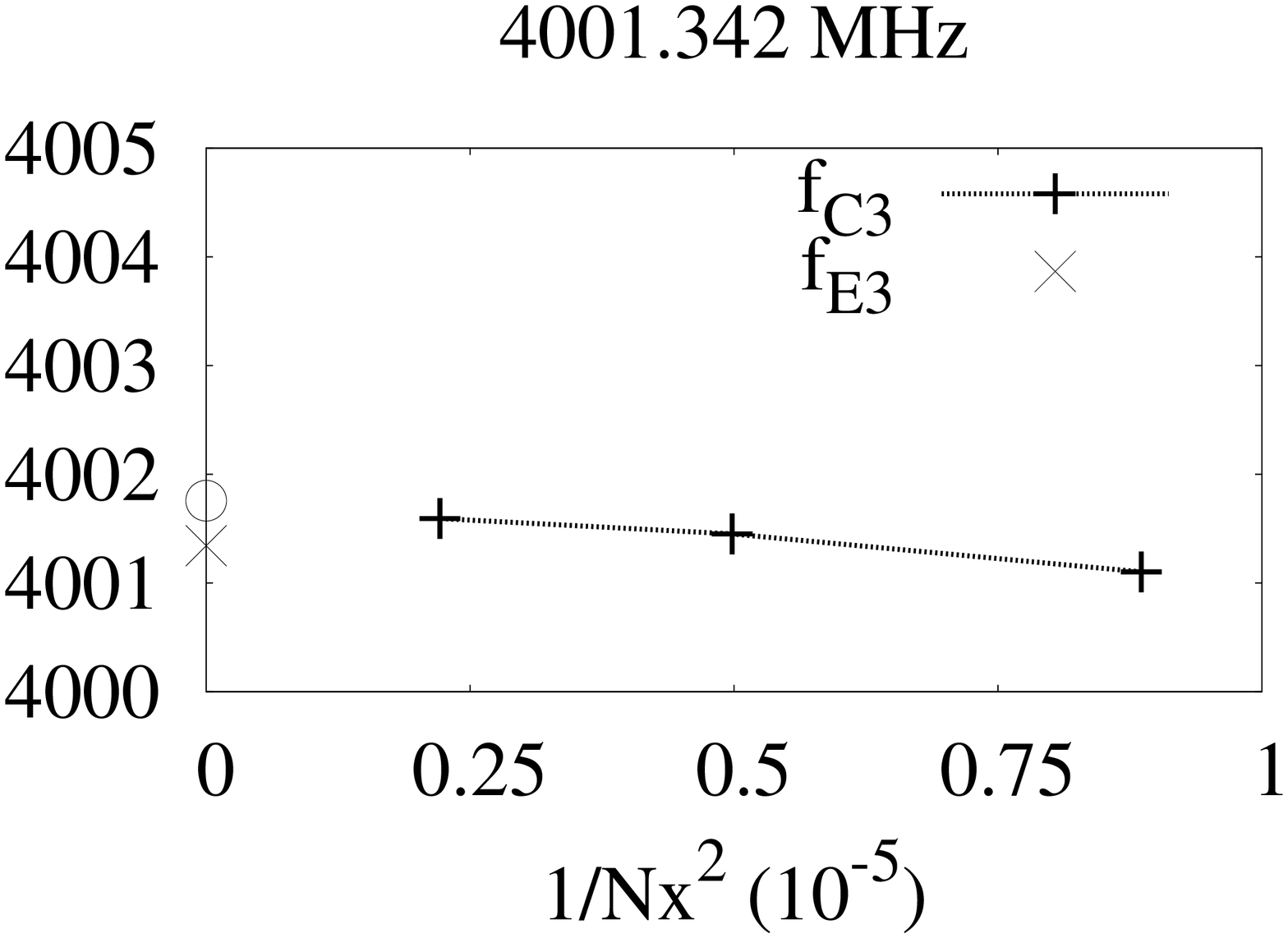,width=72mm}} % gnuplot/a15cav3conv3.pdf
\hspace{-10mm}
\subfigure{\epsfig{figure=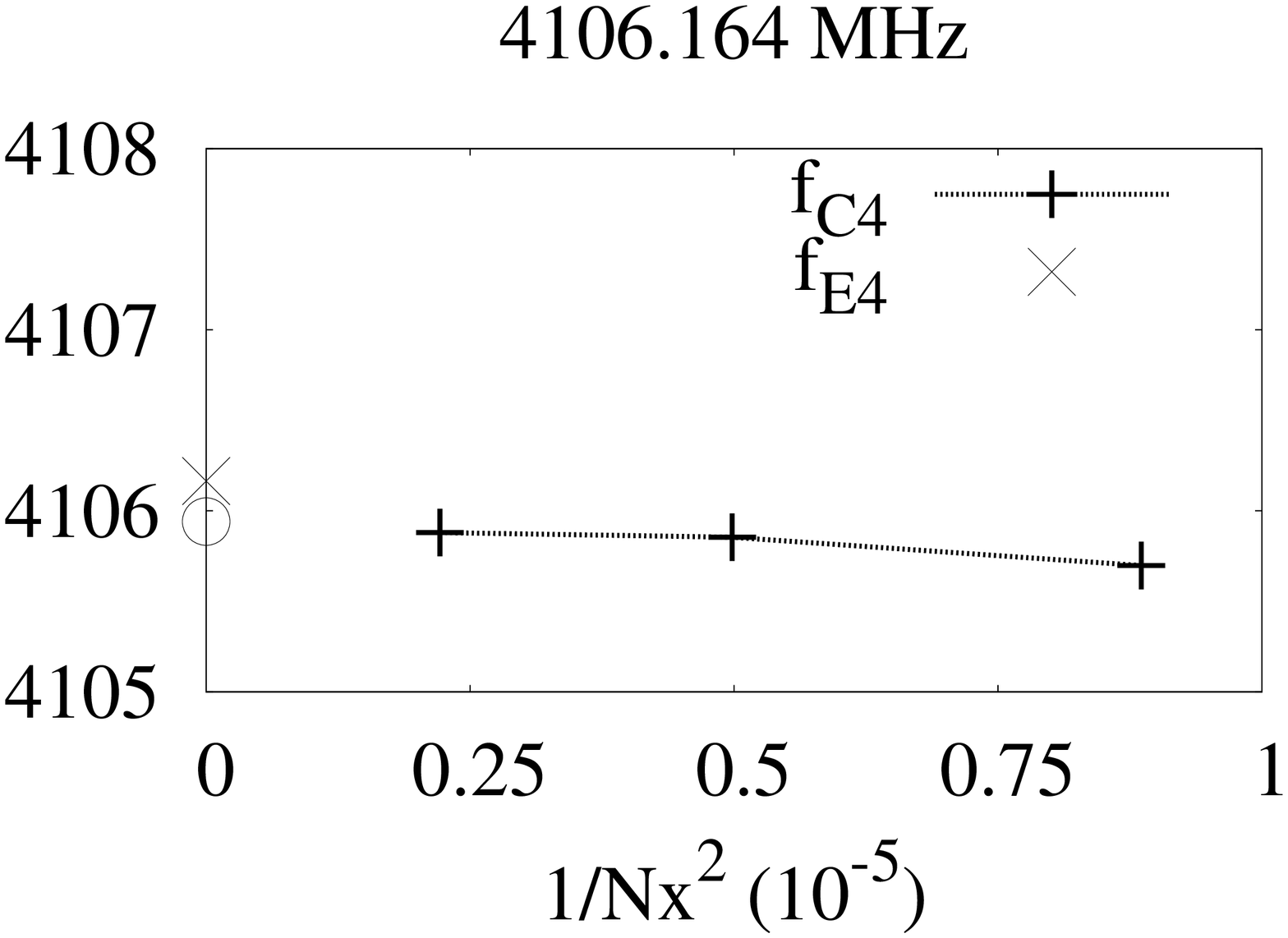,width=72mm}} % gnuplot/a15cav3conv4.pdf
\caption{Convergence plot ($\mathrm{f}_{\mathrm{C} N}$) of the frequencies of 
  the five deflecting modes. The abscissa represents the grid resolution
  and the ordinate corresponds to the frequency value in MHz.  Each
  plot has on the ordinate the experimental values (f$_{\mathrm{E}N}$) from Table
  \ref{tab:deflectFreqs} for reference purposes as well as the
  Richardson extrapolated values as unfilled circles.  The geometrical
  measurements used for the A15 cavity were those found in Reference
  \cite{Bellantoni:2000} but with a shifted equatorial radius to
  account for machining error.  The actual equatorial radius was
  47.165 mm versus the original 47.19 mm.}
\label{fig:a15cav3conv}
\end{center}
\end{figure}
%%%%%%%%%%%%%%%%%%%%%%%%%%%%%%%%%%%%%%%%%%%%%%%%%%%%%%%%%%%%%%%%%%%%%%%

\begin{figure}[t]
\begin{center}
\epsfig{figure=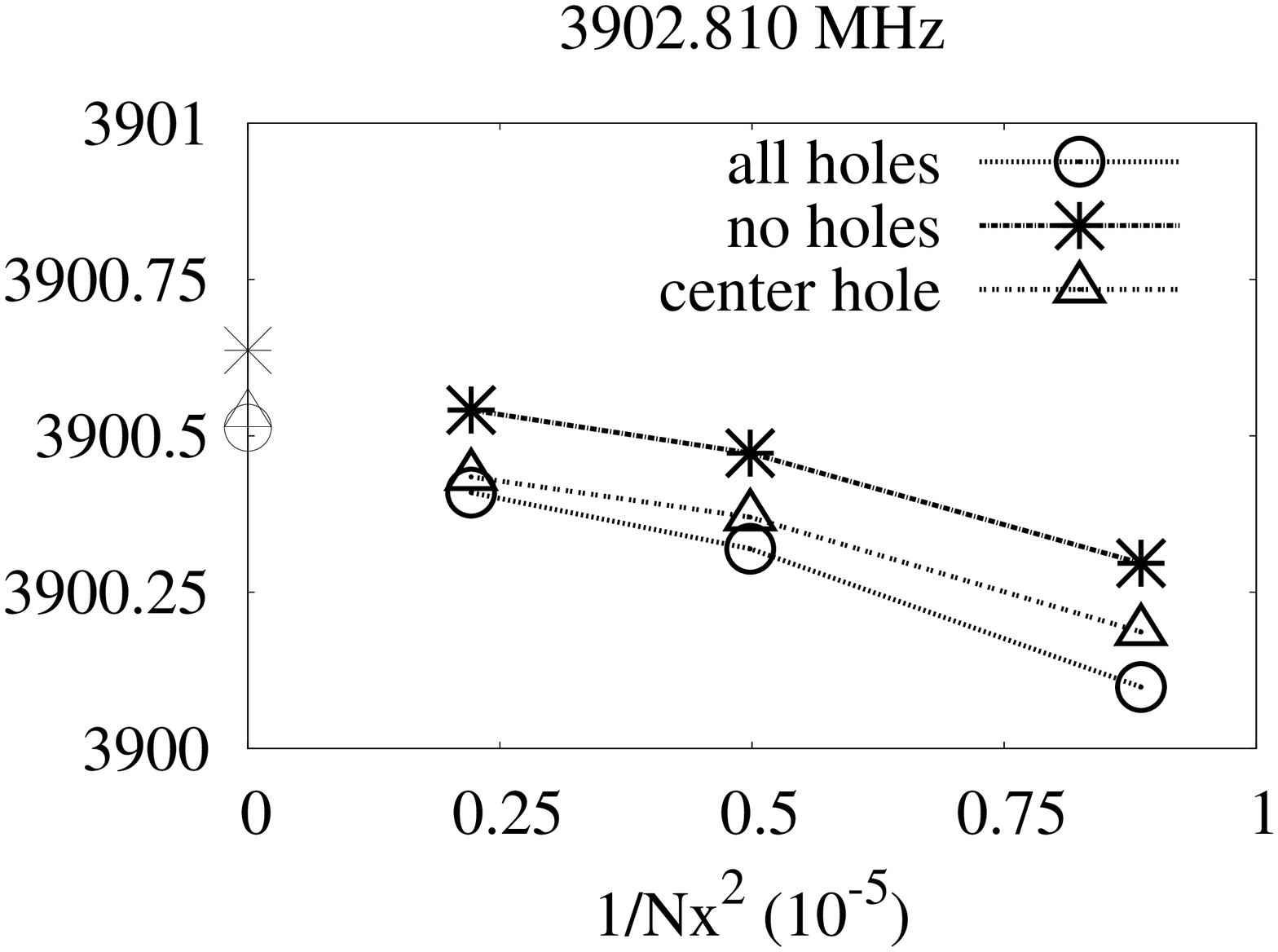,width=72mm} % gnuplot/a15cav2compare.eps
\caption{Convergence plot of the lowest frequency (the $\pi$ mode) of the A15 cavity for the cases of
all endplate holes included, no holes included, and only the center hole included.  The values on the
ordinate are the Richardson extrapolated values.  The maximum difference of values is 125 kHz using
the Richardson extrapolated values.  Note that removing the small off-center holes only shifts the 
frequencies by tens of kHz. }
\label{fig:fig5}
\end{center}
\end{figure}

As noted earlier, we believed the computed frequencies had errors far below 1 MHz, since the software had been previously verified
by comparison against analytical results (e.g, for spherical and pillbox cavities); that is, we believed the software was
correctly solving the idealized problem (Maxwell's equations in a uniform, lossless medium surrounded by a perfect conductor).
However, our cavity differed qualitatively from cavities with analytical solutions because our cavity had small holes in its
endplates (observed in Fig.~\ref{fig:a15scheme}) that were not well resolved by the computational mesh.  Completely removing all
holes from the simulation shifted resonances by less than 125 kHz, and just removing the small off-center holes only shifts the
resonances by 10s of kHz.  See Fig.~\ref{fig:fig5}.  Therefore, any error due to poor resolution of all of holes would very likely be less 
than 125 kHz, not enough to explain the differences, which are greater than 1 MHz.

%
% 3900.513 (all holes)
% 3900.5334 (only center) 
% 3900.6367 (no holes)
%
% Max Diff = 0.1237 Mhz = 123.7 kHz
%

%%%%%%%%%%%%%%%%%%%%%%%%%%%%%%%%%%%%%%%%%%%%%%%%%%%%%%%%%%%%%%%%%%%%%%%
%  FIGURE - Commented out for journal and submitted separately
%   File : a15cav3conv1.eps a15cav3conv2.eps
%          a15cav3conv3.eps a15cav3conv4.eps
%%%%%%%%%%%%%%%%%%%%%%%%%%%%%%%%%%%%%%%%%%%%%%%%%%%%%%%%%%%%%%%%%%%%%%%

At the top of our list of missing physics were material properties that vary from the ideal or cannot be precisely
determined: for example, variation in the dielectric constant of air, or skin-depth effects in the cavity walls.  A simple
estimate shows that the modification due to the finite skin depth is too small to account for the discrepancy by a few orders of
magnitude.  As for the atmosphere, Ref.~\cite{Bellantoni:2000} corrected for this; the remaining uncertainty in the dielectric
constant of air would cause only a $\pm 95$ kHz variation, which is too small to explain the differences between computation and
experiment.

Next, we considered possible differences between the cavity specifications (used in the simulation) and the actual cavity
measurements, e.g., due to machining error.  Burt {\em et al.} in Ref.~\cite{Burt:2007} had recently investigated the sensitivity
of frequency to various cavity dimensions like equatorial radius,iris radius, and cell half length for a similar superconducting
dipole cavity operating at 3.9 GHz.  The sensitivity of the frequency with respect to the equatorial radius was found to be
$-80.6$ MHz/mm.  The sensitivities for iris radius and cell half length were measured to be $-25.8$ MHz/mm and $17.4$ MHz/mm,
respectively.  A typical machining error of 1 mil or $0.025$ mm could therefore explain the observed discrepancy.

With this in mind, we computed sensitivities to equatorial radius ($\delta_r \equiv \frac{\partial\omega}{\partial r}$) and cell length ($\delta_z \equiv \frac{\partial\omega}{\partial z}$) by simulating cavities with 
slightly different equatorial radius $r$ and then with slightly different cell length $z$ (with all cells the same).  The resulting sensitivities 
(for $f_0$) are $\delta_r = -76.0$ MHz/mm, and $\delta_z = 13.4$ MHz/mm, affirming the possibility that typical machining tolerances explain 
frequency differences greater than 1 MHz.

Consequently, a careful re-measurement of the A15 cavity was made at Fermilab.  The Cordex, a coordinate measurement
machine, showed that the average equatorial radius in the fabricated cavity differed by about $0.025$ mm from the specified
value \cite{Bellantoni:2000}.  The average iris radius was only off by about $0.003$ mm, which could be ignored given
the small sensitivity of frequency to iris radius.  The cell length was measured using calipers with multiple methodologies.  From these 
measurements, we determined that the equatorial radius was $47.165 \pm 0.007$ mm (specification: $47.19$ mm), and the cell length of
$38.412 \pm 0.025$ mm (specification: $38.4$ mm).

We then simulated the cavity, changing the radius to the average measured radius, resulting in the frequencies shown in
Fig.~\ref{fig:a15cav3conv}, which are much closer to the measured values.  For comparison with Table \ref{tab:deflectFreqs}, we include 
the Richardson extrapolated values for the smaller equatorial radius with the original experimental values in Table \ref{tab:deflectFreqs2}.
Finally, for the mode $f_0$, we corrected the frequency to that of a different-length cavity using the calculated sensitivity $\delta_z$, yielding a 
computed frequency $f_0 =
3.9025$ GHz, which is only 310 kHz lower than the measured frequency, $3.90281$ GHz (differing by less than a 
part in $10^4$).  The differences between specified and fabricated cavities appear to explain the differences between computational
and measured frequencies.

\begin{table}[tb]
\begin{center}
\begin{tabular}{|c|c|c|c|c|c|} \hline
 & $f_0$ & $f_1$ & $f_2$ & $f_3$ & $f_4$ \\
\hline
Exact & 3902.810 & 3910.404 & 3939.336 & 4001.342 & 4106.164 \\
\hline 
Computed & 3902.514 & 3910.509 & 3939.155 & 4001.757 & 4105.940  \\
\hline
\end{tabular}
\end{center} \caption{\label{tab:deflectFreqs2} The first line is the set of frequencies
(in MHz) of the five deflecting modes for the A15 cavity.   The second line is the
set of computations using the altered specifications with an equatorial radius
reduced to 47.165 mm from 47.19 mm.}
\end{table}

Having understood the major source of the discrepancy, we now review the known experimental uncertainties and compare them to
the estimated computational error for $f_0$.  Uncertainty in the dielectric constant of air leads to an uncertainty (in the
corrected frequency for the cavity in vacuum) of $\pm 95$ kHz.  Uncertainty in cavity radius leads to a frequency uncertainty of
$\pm 530$ kHz; uncertainty in the length leads to $\pm 340$ kHz.  Adding these in quadrature yields an experimental uncertainty of
$640$ kHz.  We believe the computational error is negligible in comparison, on the order of 10--40 kHz.  The discrepancy between
computation and measurement was $310$ kHz, which falls well within the measurement uncertainty.  Therefore, the simulation
results agree with experiment to a part in $10^4$; experimental limitations prevent more precise validation.

\begin{figure}[bh]
\begin{center}
\epsfig{figure=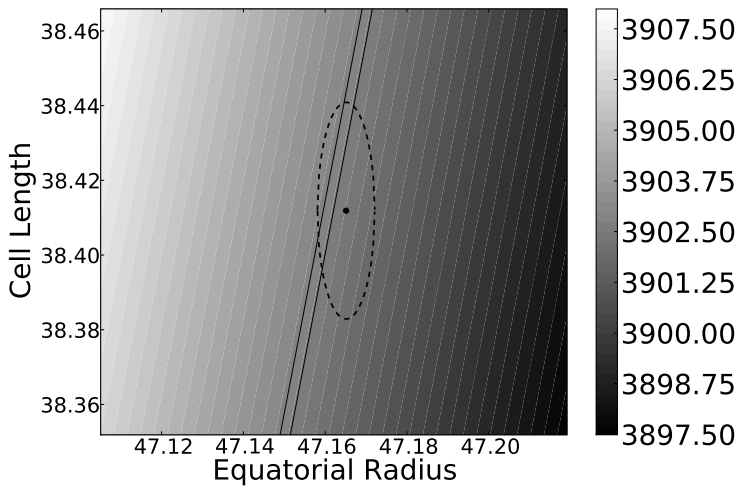,width=100mm} %gnuplot/varyF0contour_bw.eps {.png.eps}
\caption{Plot of the dependency of frequency on the equatorial
  radius and the cell length produced by calculating frequency
  gradients dependent on varying both parameters and extracting
  frequencies.  Also shown is the range of experimental values around
  the equatorial radius and cell length value, corresponding to
  $3902.810$ MHz with an experimental error of $\pm 95$ kHz, and the
  range of computed values given dimensional uncertainties.
\label{fig:contourPlot}}
\end{center}
\end{figure}

Fig.~\ref{fig:contourPlot} provides a visual analysis to understand whether this difference is significant.  Overall this
is a plot of contours of frequency in the space of equatorial radius (abscissa) and half length (ordinate).  The band between
the two black lines corresponds to the measured frequency ($3.90281$ GHz) with width given by the computational and
atmospheric correction uncertainties ($\pm 124$ kHz).  The central dot is the best computational value for the best measured
values of equatorial radius and half length.  Finally, the ellipse is one with elliptical radii given by the uncertainties
in equatorial radius and half length.  The fact that the ellipse overlaps the band indicates that the discrepancy between computed
and measured frequency is within the sum of the various uncertainties.  This figure also shows that the dominant
contributor to uncertainty in the comparison is due to the lack of precise knowledge of the dimensions of the cavity.

%%%%%%%%%%%%%%%%%%%%%%%%%%%%%%%%%%%%%%%%%%%%%%%%%%%%%%%%%%%%%%%%%%%%%%%
%  FIGURE - Commented out for journal and submitted separately
%   File : varyF0contour.eps
%%%%%%%%%%%%%%%%%%%%%%%%%%%%%%%%%%%%%%%%%%%%%%%%%%%%%%%%%%%%%%%%%%%%%%%

\section{Mode Extraction}

As has been noted previously, an important property of this method is the ease with which modes can be extracted.  To this
end, we now consider the extraction of the five deflecting modes (i.e., the eigenvectors of the operator).  Furthermore, for the
$\pi$ deflection mode, we present a complete plot consisting of electric field lines, magnetic field lines, and the magnetic
field magnitude on the cavity surface.

To begin, we revisit the algorithm for extracting modes and present the procedure with respect to the VORPAL software
\cite{Nieter:2004}, a finite-difference time-domain electromagnetic particle-in-cell code turned into an eigensolver
for this work.  VORPAL is a massively parallel code that stores data in the Hierarchical Data Format (HDF5) following the
VizSchema \cite{vizschema} conventions.  This permits dumping of fields (magnetic or electric, for example) at arbitrary times for
viewing with tools like VisIt \cite{visit}.

Once the generalized eigenvalue problem $S^{\dagger} R \mathbf{a}_m = \lambda_m S^{\dagger} S
\mathbf{a}_m$ is solved, the modes can be assembled according to
\begin{equation}
\label{eq:modeReconstruct}
 {\bf v}_m = \sum_{\ell=1}^L a_{m,\ell} {\bf s}_\ell.
\end{equation}
While ${\bf a}_m$ is the eigenvector of $S^{\dagger} R
\mathbf{a}_m = \lambda_m S^{\dagger} S \mathbf{a}_m$, ${\bf v}_m$ is the eigenvector for the discrete operator $\curl \curl$. In a VORPAL
simulation, the vectors ${\bf s}_{\ell}$ correspond to dumps of the magnetic (or electric) field at given times.  A requirement is that all
temporal dumps correspond entirely to the magnetic (or electric) field when constructing a magnetic (or electric) field mode.

%%%%%%%%%%%%%%%%%%%%%%%%%%%%%%%%%%%%%%%%%%%%%%%%%%%%%%%%%%%%%%%%%%%%%%%
%  FIGURE - Commented out for journal and submitted separately
%   File : bmode0.eps bmode1.eps bmode2.eps bmode3.eps bmode4.eps
\begin{figure}[bht]
\begin{center}
\subfigure[~3902.810 MHz]{\epsfig{figure=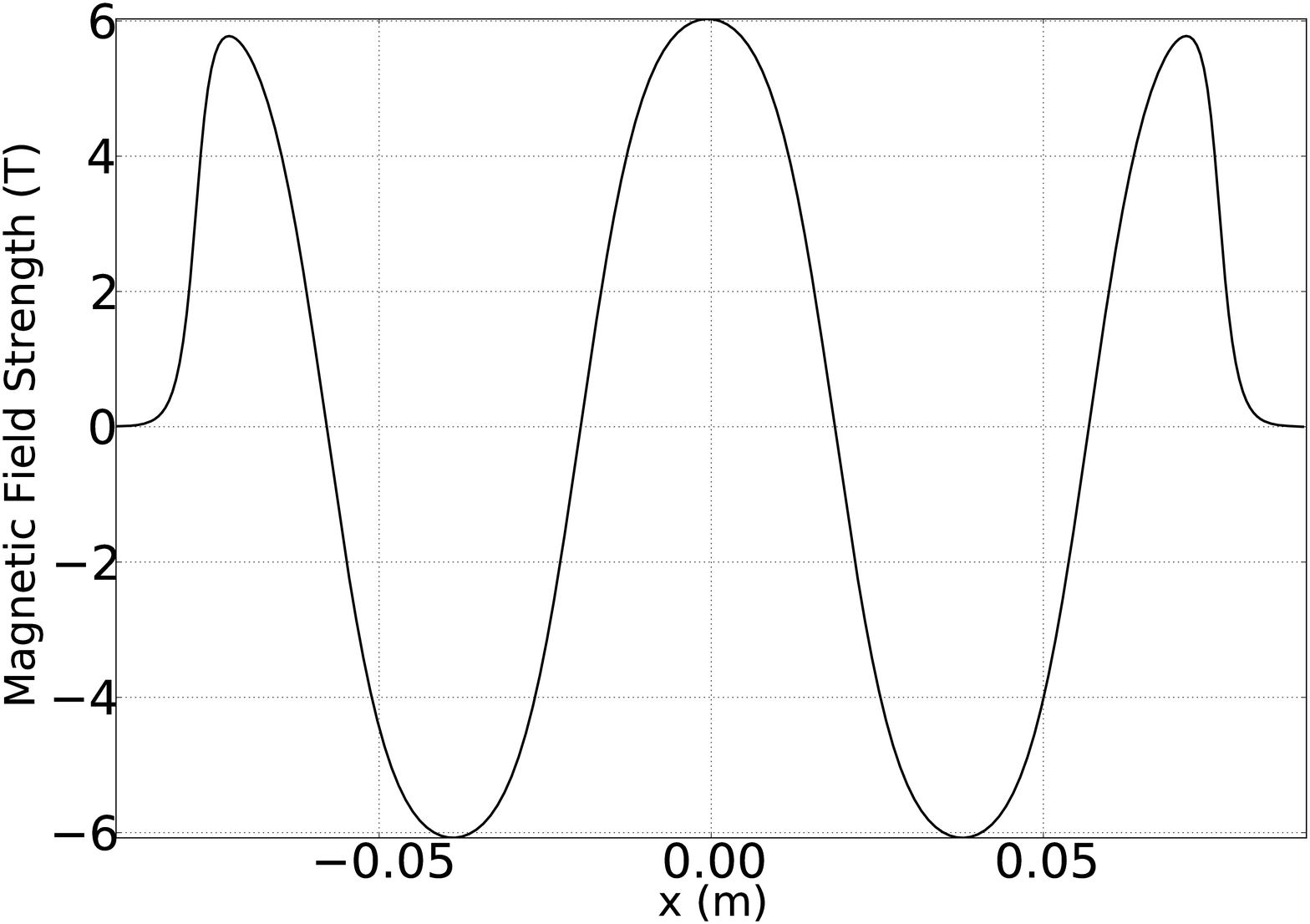,width=75mm} } % Visit/mode0.eps
\hspace{5mm}
\subfigure[~3910.404 MHz]{\epsfig{figure=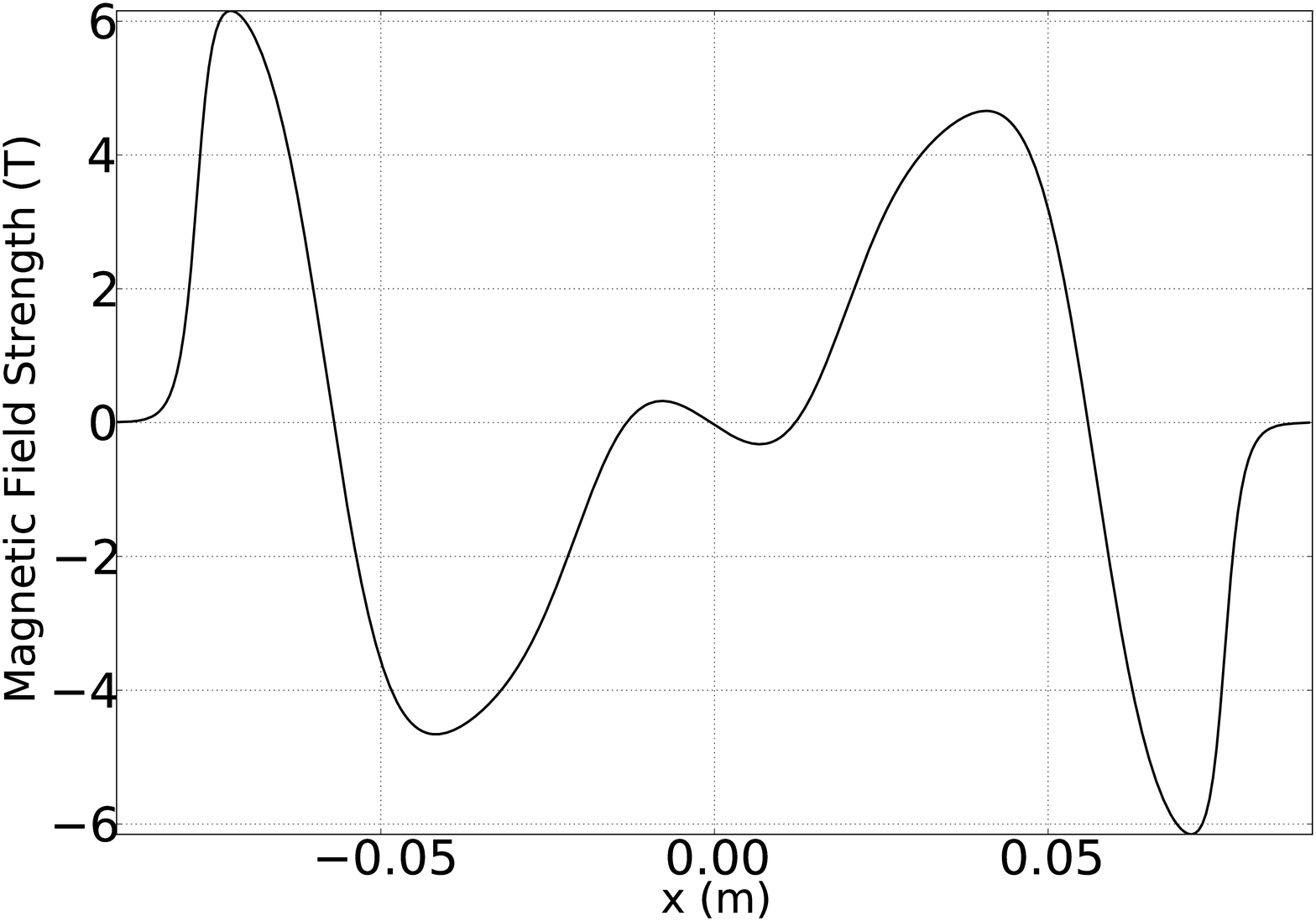,width=75mm} } % Visit/mode1.eps
\subfigure[~3939.336 MHz]{\epsfig{figure=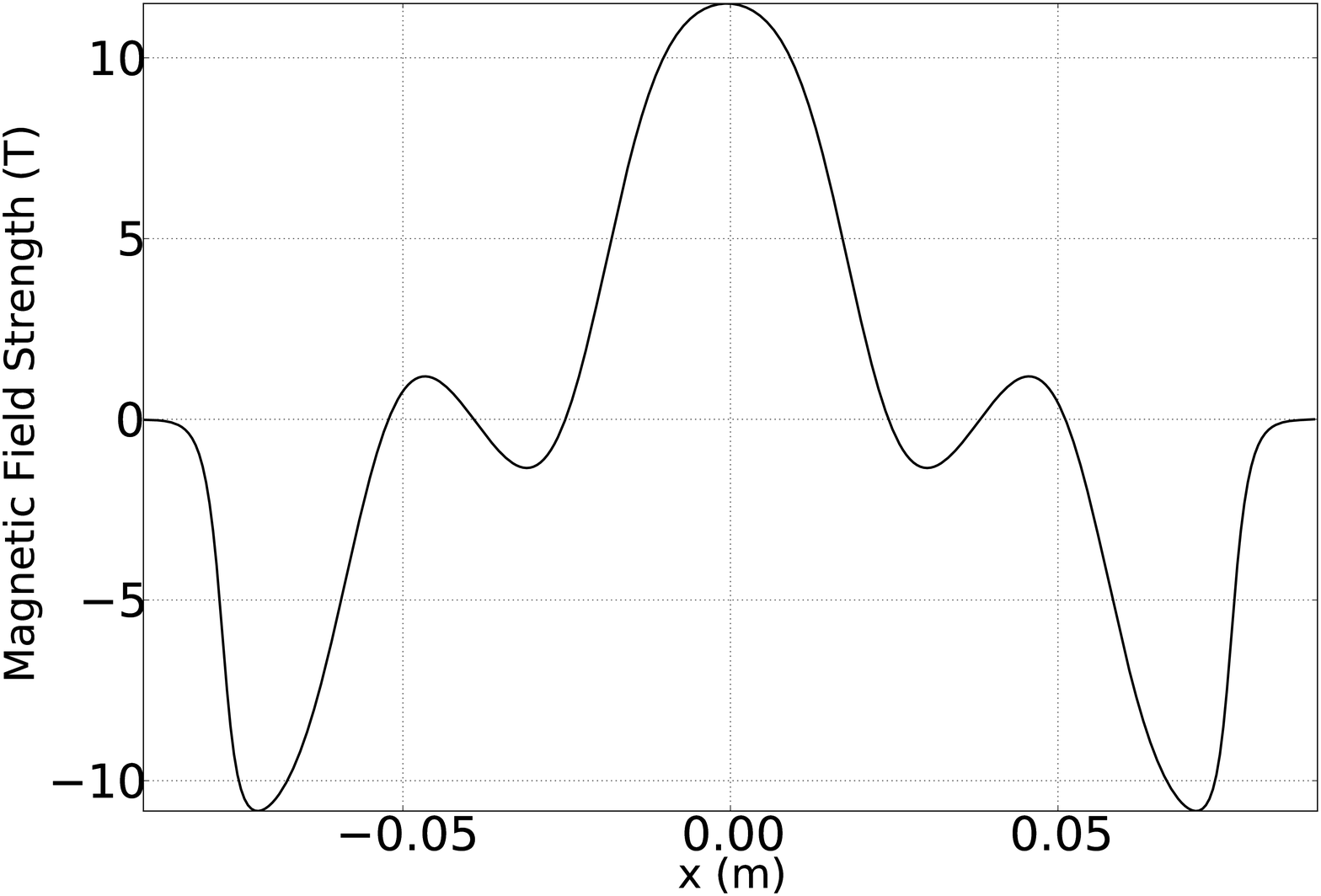,width=75mm} } % Visit/mode2.eps
\hspace{5mm}
\subfigure[~4001.342 MHz]{\epsfig{figure=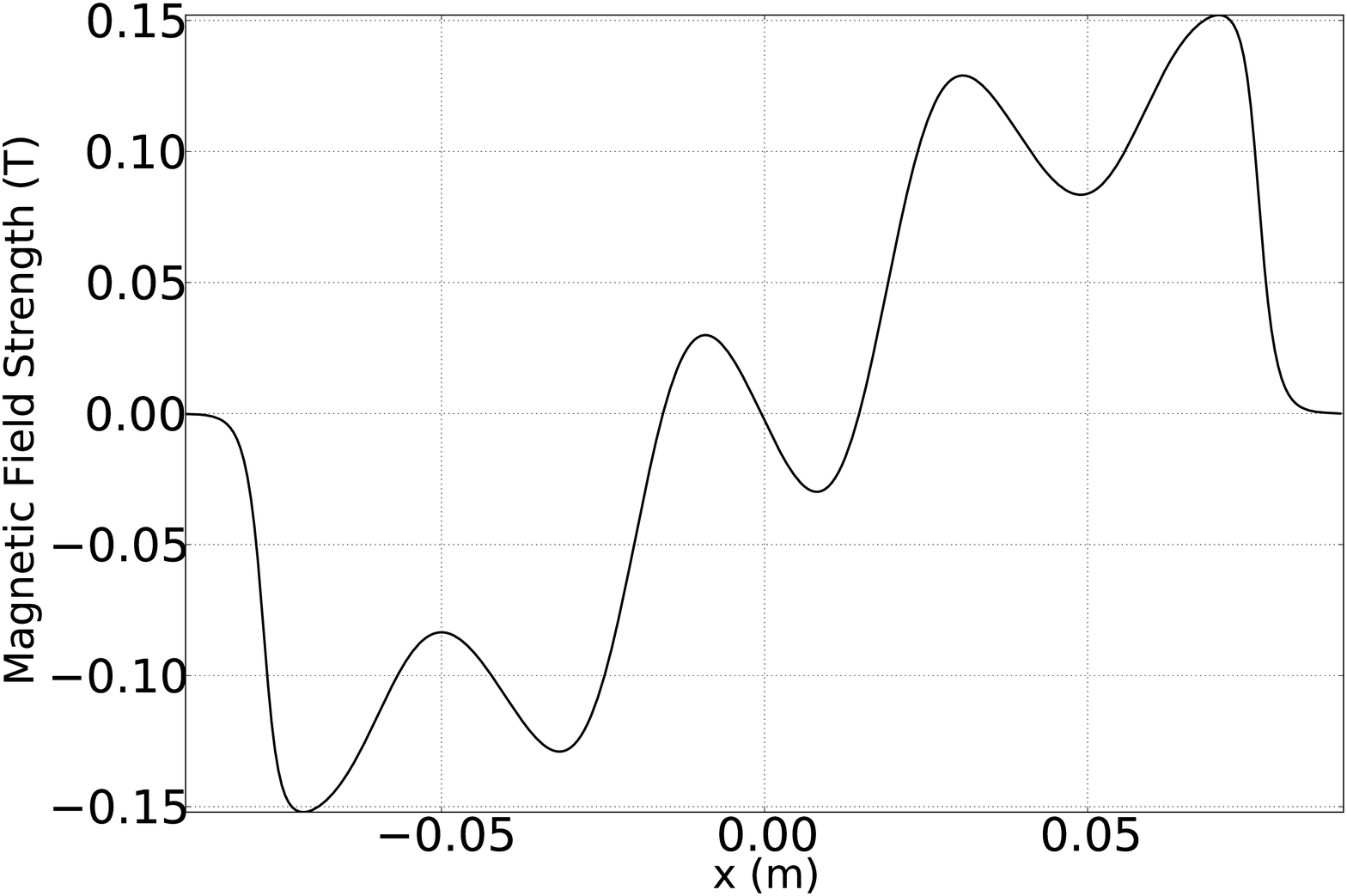,width=75mm} } % Visit/mode3.eps
\subfigure[~4106.164 MHz]{\epsfig{figure=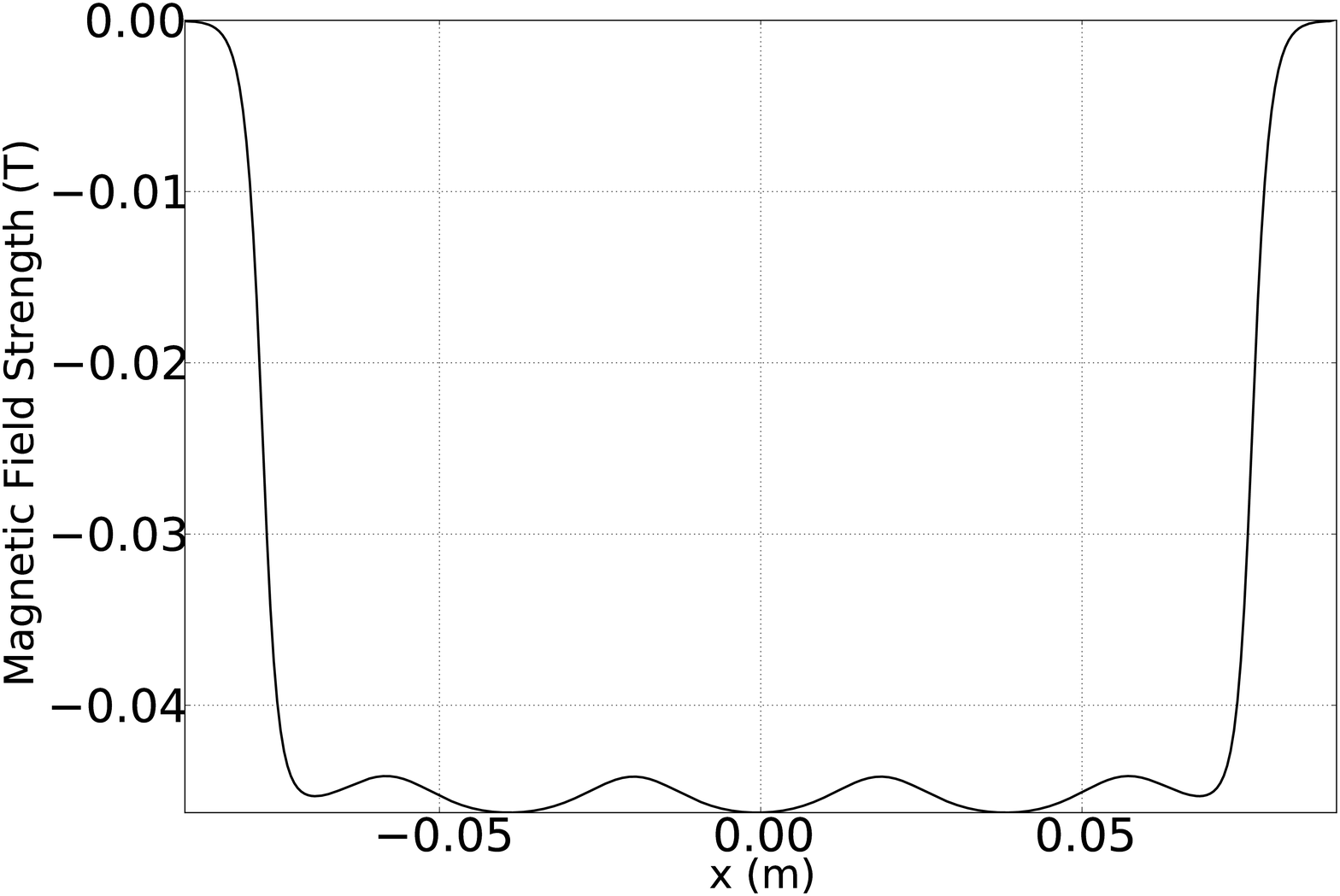,width=75mm} } % Visit/mode4.eps
\caption{ Lineout of z-component of the magnetic field ($B_z$) at r=0 for the five
  deflecting (TM$_{110}$) modes of the A15 cavity.  Simulations were
  ran at a resolution of $1.07  \times 1.3 \times 1.3 \; \mathrm{mm}^3$
  resulting in approximatley 1.05 million grid points.  Each mode
  reconstruction, which consisted of constructing an hdf5 file as a
  linear combination of 20 hdf5 files, took less than a minute to
  perform.
 \label{fig:fourModes}}
\end{center}
\end{figure}
%%%%%%%%%%%%%%%%%%%%%%%%%%%%%%%%%%%%%%%%%%%%%%%%%%%%%%%%%%%%%%%%%%%%%%%

As each ${\bf s}_{\ell}$ is represented by an HDF5 file, we must take linear combinations of multiple HDF5 files to assemble the
modes.  Python and its PyTables package are used to perform the assembly on a single processor.  For approximately 1.05 million
grid points it took less than a minute to perform a single mode assembly. In Fig.~\ref{fig:fourModes} we see the magnetic field
of the five deflecting modes on a grid with a resolution of $1.07
\times 1.3 \times 1.3 \; \mathrm{mm}^3$.  The $\pi$ mode exhibits expected behavior.  We also 
see the behavior exhibited by remaining deflecting modes.

\begin{figure}[t]
\begin{center}
\epsfig{figure=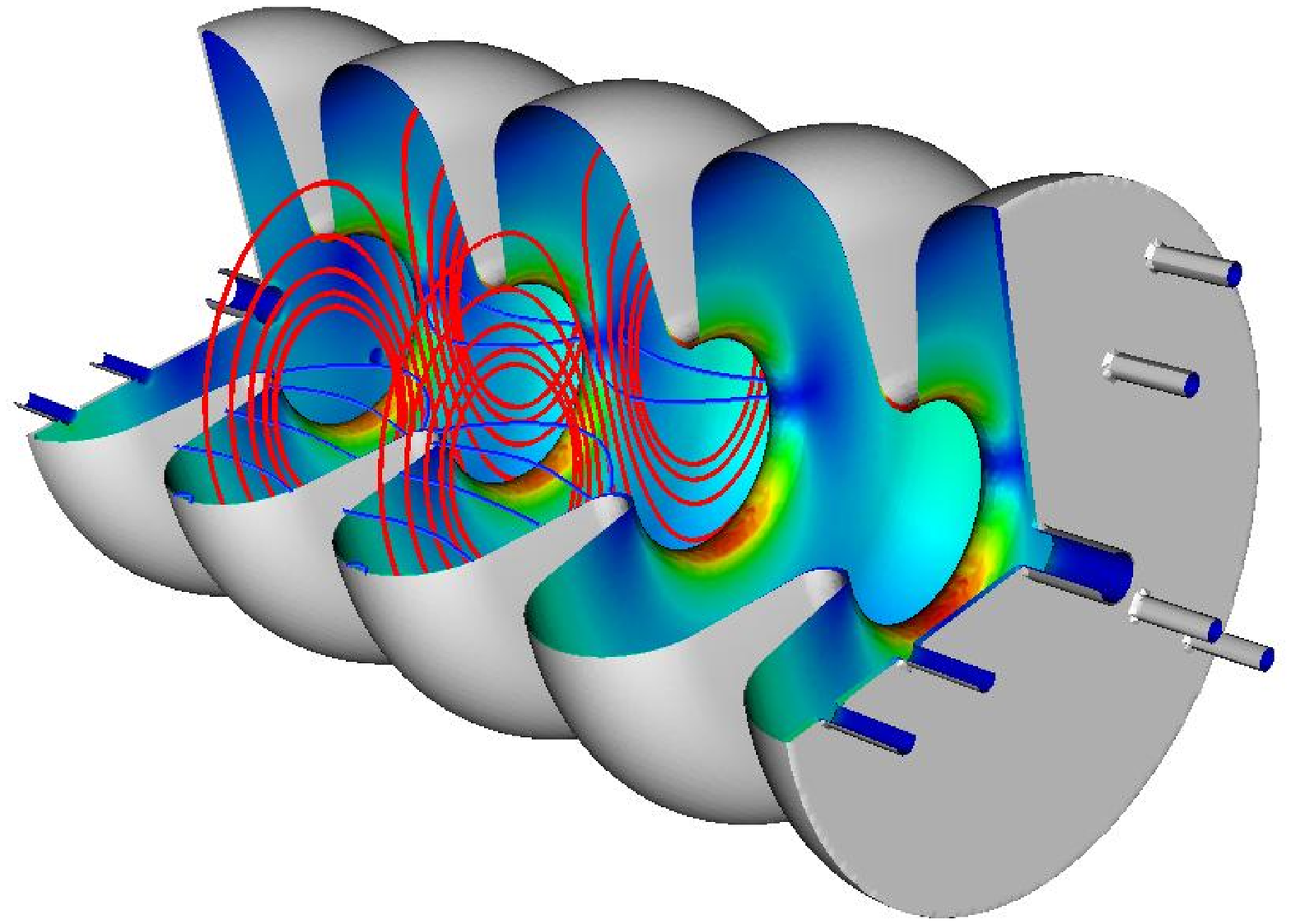,width=100mm}  % Visit/sciDacImage0000.pdf {eps,png}
\end{center}
\caption{(Color online) Full representation of $\pi$ mode for the A15 cavity
  illustrating the electric (blue/darker) and the magnetic (red/lighter) field lines
  and the magnitude of the magnetic field painted on the inside of the
  cavity.  All modes were constructed from simulations with
  post-processing to permit representation within the cavity.  Image
  used with permission of the DOE SciDAC Office. \label{fig:sciDac}}
\end{figure}

In Fig.~\ref{fig:sciDac} we have assembled both the electric field and the magnetic field and also used a Python script to obtain
the magnetic field magnitude on the cavity surface for the $\pi$ mode.   The magnetic field magnitude shows areas on the cavity that may 
result in quenching, which is critical information that can be used by engineers and physicists considering cavity design.

%%%%%%%%%%%%%%%%%%%%%%%%%%%%%%%%%%%%%%%%%%%%%%%%%%%%%%%%%%%%%%%%%%%%%%%
%  FIGURE - Commented out for journal and submitted separately
%   File : sciDacImage0000.eps
%%%%%%%%%%%%%%%%%%%%%%%%%%%%%%%%%%%%%%%%%%%%%%%%%%%%%%%%%%%%%%%%%%%%%%%

\section{Conclusion}
\label{sec:conc}

The frequency and mode extraction algorithm proposed in Ref.~\cite{WernerCary:2008} in combination with the VORPAL
computational framework \cite{Nieter:2004} has been shown to efficiently produce frequencies and modes for a realistic
accelerator cavity, in this case a stack of four unpolarized dumbbells assembled in the form of a cavity that was designed in
the past for an experimental program at Fermilab.  We also showed that this method can extract the modes of the cavity.  We
estimated that this method, along with Richardson extrapolation, was able to compute frequencies to better than a part in $10^5$.

We further outlined the validation steps needed for determining whether this method is accurate.  Given that the computational
software had been verified, we proceeded through the validation process of considering errors in the measurement, failure to
include relevant physics, and differences between the experiemental and computational geometrical models.  Ultimately,
we found that the computational frequency and the experimentally measured frequency were consistent to within the uncertainties
(roughly $1 \times 10^{-4}$), with the largest contributor to uncertainty resulting from imprecisely knowing the dimensions
(particularly the equatorial radius) of the cavity.  This is important, as it provides some guidance for future comparisons of
calculations with measurements -- namely, one must understand the dimensional uncertainties as a first step.

\section{Acknowledgments}
\label{sec:ack}

This work was supported by the U.S. Department of
Energy grants DE-FG02-04ER41317, DE-FC02-07ER41499,
and DE-AC02-07CH11359.

We also thank the VORPAL team, T.~Austin, G.~I.~Bell,
D.~L.~Bruhwiler, R.~S.~Busby, M.~Carey, J.~Carlsson, J.~R.~Cary,
Y.~Choi, B.~M.~Cowan, D.~A.~Dimitrov, A.~Hakim, J.~Loverich,
S.~Mahalingam, P.~Messmer, P.~J.~Mullowney, C.~Nieter, K.~Paul,
C.~Roark, S.~W.~Sides, N.~D.~Sizemore, D.~N.~Smithe,
P.~H.~Stoltz, S.~A.~Veitzer, D.~J.~Wade-Stein, G.~R.~Werner,
M.~Wrobel, N.~Xiang, C.~D.~Zhou.

\bibliography{validate_freq_extract}

\begin{thebibliography}{17}
\expandafter\ifx\csname natexlab\endcsname\relax\def\natexlab#1{#1}\fi
\expandafter\ifx\csname bibnamefont\endcsname\relax
  \def\bibnamefont#1{#1}\fi
\expandafter\ifx\csname bibfnamefont\endcsname\relax
  \def\bibfnamefont#1{#1}\fi
\expandafter\ifx\csname citenamefont\endcsname\relax
  \def\citenamefont#1{#1}\fi
\expandafter\ifx\csname url\endcsname\relax
  \def\url#1{\texttt{#1}}\fi
\expandafter\ifx\csname urlprefix\endcsname\relax\def\urlprefix{URL }\fi
\providecommand{\bibinfo}[2]{#2}
\providecommand{\eprint}[2][]{\url{#2}}

\bibitem[{\citenamefont{Werner and Cary}(2008)}]{WernerCary:2008}
\bibinfo{author}{\bibfnamefont{G.}~\bibnamefont{Werner}} \bibnamefont{and}
  \bibinfo{author}{\bibfnamefont{J.}~\bibnamefont{Cary}}, \bibinfo{journal}{J.
  Comput. Phys.} \textbf{\bibinfo{volume}{227}}, \bibinfo{pages}{5200}
  (\bibinfo{year}{2008}).

\bibitem[{\citenamefont{Nieter and Cary}(2004)}]{Nieter:2004}
\bibinfo{author}{\bibfnamefont{C.}~\bibnamefont{Nieter}} \bibnamefont{and}
  \bibinfo{author}{\bibfnamefont{J.}~\bibnamefont{Cary}}, \bibinfo{journal}{J.
  Comput. Phys.} \textbf{\bibinfo{volume}{196}}, \bibinfo{pages}{448}
  (\bibinfo{year}{2004}).

\bibitem[{\citenamefont{{M. Clemens et al.}}(1999)}]{MAFIA:1999}
\bibinfo{author}{\bibnamefont{{M. Clemens et al.}}}, in
  \emph{\bibinfo{booktitle}{Proceedings of the 1999 Int.~Conf.~on
  Comp.~Electromagnetics and its applications}} (\bibinfo{year}{1999}), pp.
  \bibinfo{pages}{565--568}.

\bibitem[{\citenamefont{McAshan and Wanzenberg}()}]{McAshan:2001}
\bibinfo{author}{\bibfnamefont{M.}~\bibnamefont{McAshan}} \bibnamefont{and}
  \bibinfo{author}{\bibfnamefont{R.}~\bibnamefont{Wanzenberg}},
  \bibinfo{note}{{F}ermi {N}ational {A}ccelerator {L}ab {T}echnical {N}ote
  TM-2144}.

\bibitem[{\citenamefont{Bellantoni}()}]{Bellantoni:2000}
\bibinfo{author}{\bibfnamefont{L.}~\bibnamefont{Bellantoni}},
  \bibinfo{note}{unpublished note}.

\bibitem[{\citenamefont{Neuhauser}(1990)}]{Neuhauser:1990}
\bibinfo{author}{\bibfnamefont{D.}~\bibnamefont{Neuhauser}},
  \bibinfo{journal}{J. Chem. Phys.} \textbf{\bibinfo{volume}{93}},
  \bibinfo{pages}{2611} (\bibinfo{year}{1990}).

\bibitem[{\citenamefont{Neuhauser}(1994)}]{Neuhauser:1994}
\bibinfo{author}{\bibfnamefont{D.}~\bibnamefont{Neuhauser}},
  \bibinfo{journal}{J. Chem. Phys.} \textbf{\bibinfo{volume}{100}},
  \bibinfo{pages}{5076} (\bibinfo{year}{1994}).

\bibitem[{\citenamefont{Wall and Neuhauser}(1995)}]{Wall:1995}
\bibinfo{author}{\bibfnamefont{M.}~\bibnamefont{Wall}} \bibnamefont{and}
  \bibinfo{author}{\bibfnamefont{D.}~\bibnamefont{Neuhauser}},
  \bibinfo{journal}{J. Chem. Phys.} \textbf{\bibinfo{volume}{102}},
  \bibinfo{pages}{8011} (\bibinfo{year}{1995}).

\bibitem[{\citenamefont{Mandelshtam and Taylor}(1997)}]{Mandelshtam:1997}
\bibinfo{author}{\bibfnamefont{V.}~\bibnamefont{Mandelshtam}} \bibnamefont{and}
  \bibinfo{author}{\bibfnamefont{H.}~\bibnamefont{Taylor}},
  \bibinfo{journal}{J. Chem. Phys.} \textbf{\bibinfo{volume}{107}},
  \bibinfo{pages}{6756} (\bibinfo{year}{1997}).

\bibitem[{\citenamefont{Mandelshtam}(2003)}]{Mandelshtam:2003}
\bibinfo{author}{\bibfnamefont{V.}~\bibnamefont{Mandelshtam}},
  \bibinfo{journal}{J. Theor. Comput. Chem.} \textbf{\bibinfo{volume}{2}},
  \bibinfo{pages}{497} (\bibinfo{year}{2003}).

\bibitem[{\citenamefont{Yee}(1966)}]{Yee:1966}
\bibinfo{author}{\bibfnamefont{K.~S.} \bibnamefont{Yee}},
  \bibinfo{journal}{IEEE Trans. Antennas Propag.}
  \textbf{\bibinfo{volume}{14}}, \bibinfo{pages}{302} (\bibinfo{year}{1966}).

\bibitem[{\citenamefont{Bellantoni et~al.}(2001)\citenamefont{Bellantoni,
  Edwards, McAshan, and Wanzenberg}}]{Bellantoni:2001}
\bibinfo{author}{\bibfnamefont{L.}~\bibnamefont{Bellantoni}},
  \bibinfo{author}{\bibfnamefont{H.}~\bibnamefont{Edwards}},
  \bibinfo{author}{\bibfnamefont{M.}~\bibnamefont{McAshan}}, \bibnamefont{and}
  \bibinfo{author}{\bibfnamefont{R.}~\bibnamefont{Wanzenberg}}, in
  \emph{\bibinfo{booktitle}{Conference: Particle Accelerator Conference,
  Chicago, IL (US), 06/18/2001--06/22/2001}} (\bibinfo{year}{2001}).

\bibitem[{\citenamefont{Dey and Mittra}(1997)}]{dey1997lcf}
\bibinfo{author}{\bibfnamefont{S.}~\bibnamefont{Dey}} \bibnamefont{and}
  \bibinfo{author}{\bibfnamefont{R.}~\bibnamefont{Mittra}},
  \bibinfo{journal}{IEEE Microwave and Guided Wave Letters}
  \textbf{\bibinfo{volume}{7}}, \bibinfo{pages}{273} (\bibinfo{year}{1997}).

\bibitem[{\citenamefont{Nieter et~al.}(2009)\citenamefont{Nieter, Cary, Werner,
  Smithe, and Stoltz}}]{Nieter:2009}
\bibinfo{author}{\bibfnamefont{C.}~\bibnamefont{Nieter}},
  \bibinfo{author}{\bibfnamefont{J.}~\bibnamefont{Cary}},
  \bibinfo{author}{\bibfnamefont{G.}~\bibnamefont{Werner}},
  \bibinfo{author}{\bibfnamefont{D.}~\bibnamefont{Smithe}}, \bibnamefont{and}
  \bibinfo{author}{\bibfnamefont{P.}~\bibnamefont{Stoltz}},
  \bibinfo{journal}{J. Comput. Phys.} \textbf{\bibinfo{volume}{in review}}
  (\bibinfo{year}{2009}).

\bibitem[{\citenamefont{Burt et~al.}(2007)\citenamefont{Burt, Bellantoni, and
  Dexter}}]{Burt:2007}
\bibinfo{author}{\bibfnamefont{G.}~\bibnamefont{Burt}},
  \bibinfo{author}{\bibfnamefont{L.}~\bibnamefont{Bellantoni}},
  \bibnamefont{and} \bibinfo{author}{\bibfnamefont{A.}~\bibnamefont{Dexter}},
  \bibinfo{journal}{EUROTeV-Report-2007-003}  (\bibinfo{year}{2007}).

\bibitem[{viz()}]{vizschema}
\emph{\bibinfo{title}{{V}iz{S}chema}},
  \bibinfo{howpublished}{\url{https://ice.txcorp.com/trac/vizschema}},
  \bibinfo{note}{last accessed July 8, 2009.}

\bibitem[{vis()}]{visit}
\emph{\bibinfo{title}{{V}is{I}t}},
  \bibinfo{howpublished}{\url{http://visit.llnl.gov}}, \bibinfo{note}{last
  accessed July 8, 2009.}

\end{thebibliography}

\end{document}